\documentclass[preprintnumbers,amsmath,amssymb,floatfix,11pt,prd,onecolumn,
superscriptaddress,nofootinbib]{revtex4}
\usepackage[utf8]{inputenc}
\usepackage{latexsym,float}
\usepackage{epsfig}
\usepackage{epstopdf}
\usepackage{amsmath, hyperref}
\usepackage{amssymb}
\usepackage{graphicx}
\usepackage[T1]{fontenc}
\begin{document}
\title{Cyclic cosmology in modified gravity}
\author{Petar Pavlovic}
\email{petar.pavlovic@desy.de}\affiliation{Institut f\"{u}r Theoretische Physik,
	Universit\"{a}t Hamburg, Luruper Chaussee 149, 22761 Hamburg, Germany}

\author{Marko Sossich}
\email{marko.sossich@fer.hr}\affiliation{University of Zagreb, Faculty of Electrical Engineering and Computing, Department of Physics,
	Unska 3, 10 000 Zagreb, Croatia}

\begin{abstract}
 {\centering \bf Abstract:}   In this work we propose a new general model of eternal cyclic Universe. We start from the assumption that quantum gravity corrections can be effectively accounted by the addition of higher order curvature terms
in the Lagrangian density for gravity. It is also taken into account that coefficients associated with these curvature corrections 
will in general be dependent on a curvature regime. We therewith assume no new ingredients, such as extra dimensions, new scalar fields, phantom energy or special space-time geometries.  Evolution of the Universe in this framework is studied and general properties of each phase
of the cycle - cosmological bounce, low curvature ($\Lambda$CDM) phase, destruction of bounded systems and contracting phase - are analysed in detail. Focusing on
some simple special cases, we obtain analytical and numerical solutions for the each phase confirming our analysis. 
\end{abstract} 

\maketitle

\section{Introduction}
 More than a century after its discovery by Albert Einstein \cite{ein1, ein2}, general relativity (GR) is still one of the most successful physical theories
and currently our best empirically verified description of gravity. In its essence, general relativity rejected the old Newtonian concept of gravitational
force and replaced it with the concept of deformation of space-time structure caused by the distribution of mass-energy. Einstein's general
relativity is shown to be consistent with various empirical observations and tests - such as the light deflection, the perihelion advance 
of Mercury, the gravitational red-shift, the gravitational wave-damping in binary pulsars, different solar system measurements, and the recent
LIGO detection of GW150914 \cite{berti,will, tur,ligo}. Einstein's discovery not only fundamentally changed our understanding of space, time and gravitational interaction, but
it also opened the way for the foundation of physical cosmology, as the study of the Universe in its totality. First steps in this direction were done
by Einstein himself \cite{ein3}, followed by the important Hubble's discovery of the expanding Universe \cite{hu1,hu2}, and theoretical models proposed 
by Friedmann and Lamaitre \cite{fri} - finally leading to the establishment of the standard model of cosmology in the following decades \cite{robertson, weinberg}.
The standard model of cosmology, based on GR and assumptions of large-scale homogeneity and isotropicity, was tested
by a plethora of observations - such as supernova and microwave background measurements, consistency of the age of the Universe 
with the age of astronomical objects,
abundances of chemical elements, growth of cosmological perturbations etc.\cite{maartens}. However, the standard model of cosmology can not explain the observed astronomical and 
cosmological dynamics
without introducing new forms of matter-energy, for which there is currently still no empirical evidence. Various observations,
from type Ia supernovae \cite{perlmutter, riess}, large scale structure \cite{eisenstein} and cosmic microwave background radiation \cite{spergel}, 
confirm that the Universe is characterized by
accelerated expansion. Moreover, different observations on astronomical \cite{ost,refre,mas}, as well as cosmological scales \cite{ol,planco} in the 
framework of standard GR require that most of the matter
in the Universe consists of some unknown non-baryonic matter. The approach within the standard $\Lambda$CDM cosmological model \cite{bull}, which assumes
complete validity of Einstein's GR, was to introduce cold dark matter and dark energy as dominant contributions to the
mass-energy of the Universe in order to 
explain the "missing mass" and accelerated expansion of the Universe.
One of the big unsolved problems of the $\Lambda$CDM model is that the cosmological constant, playing the role of dark energy, needs to have an extremely small
observed value so that it can not be simply attributed to the vacuum energy \cite{carroll, wein}.\\ 
\\
However, it can as well be the case that these observations are a consequence of our incomplete knowledge of gravity, and not the assumed existence of 
still unobserved 
dark energy and dark matter. Therefore, another popular route of research is to examine the potential generalizations of Einstein's GR, which 
could explain cosmological observations
without postulating dark energy and dark mater, and also pass all the classical tests of GR. There is a large number of 
proposed modified theories of gravity
\cite{timothy, noj-odin}, including some of the more popular models such as Brans-Dicke gravity \cite{brans}, modified $f(T)$ teleparallel 
gravity \cite{cai}, Kaluza-Klein theories \cite{overduin} 
and Ho\v{r}ava-Lifshitz gravity \cite{thom}. In the last decades special attention
was given to modified theories that preserve all fundamental physical assumptions of Einstein's GR and only generalize the 
gravitational action integral. The simplest of 
these models is probably modified $f(R)$ gravity \cite{faraoni}. In this approach the Ricci scalar curvature, $R$, is replaced by 
an arbitrary function of scalar curvature, $f(R)$, 
in the gravitational Lagrangian density. 
Research on the possible modifications and extensions of Einstein's GR can be further justified by the fact that  GR can not be a 
complete theory because it does not take into account the principles of quantum physics. Since we still do not have a developed and verified theory of
quantum gravity, work on modified gravity can potentially lead to new insights regarding its principles, 
properties and mathematical form. Indeed, it was
shown that modifications of the Lagrangian density including higher order curvature invariants can lead to quantization 
of gravity \cite{stelle, vilkovisky}. \\
\\
As proven by the singularity theorems of Hawking, if standard GR is correct and if some general and usual conditions on the 
space-time and matter (energy conditions) are satisfied, then our Universe needed to have a beginning 
in the initial singularity of the Big Bang \cite{haw1,haw2,haw3}. The idea of a beginning of the Universe leads to the philosophical problems
of creation \textit{ex nihilo} (for instance it is difficult to see how starting from the persistence of pure nothingness a tendency towards the creation of
something can arise all of a sudden). Also, this initial singularity would mean the breaking of GR,
as well as loosing the possibility of a physical description of the Universe. Moreover, the Big Bang paradigm leads also to some important
physical difficulties such as the flatness problem and horizon problem, requiring some new mechanisms for the solution of
these issues. The most popular attempt of such mechanism is inflationary cosmology \cite{linde}. However it requires even more additions
to the $\Lambda$CDM model, such as specific scalar fields to support inflation. Also, all the inflationary models have to be
fine-tuned in order that the spectrum and amplitude of primordial density perturbations agree with observations.\\
\\
But, as it will be discussed later, singularity theorems are strongly dependent on the structure of the field equations in  GR.
Even some simple generalizations of the action integral for gravity can prevent the initial Big-Bang singularity to occur, and
enable a transition from expansion of the Universe to an earlier phase of contraction - a cosmological bounce. It is usually
assumed that the quantization of gravity will wipe-out the singularities existing in  classical GR \cite{bojo}. If, in a new theory of
quantum gravity, the Big-Bang singularity will not exist, it seems plausible that the described cosmological bounce needed to occur instead. 
Extending the idea of a bounce to construct a logically consistent picture of a full evolution of the Universe, naturally leads
to the paradigm of a cyclic Universe - an infinitely existing Universe, undergoing a bounce at the beginning and at the end of each cycle, as well
as the phases of radiation, matter and dark energy domination predicted by the standard $\Lambda$CDM cosmology. All of the presented
arguments therefore naturally establish a connection between modified gravity and cyclic cosmology. \\
\\
In this work our aim will be to propose a new model of cyclic cosmology based on modified gravity. The underlaying idea is that standard GR is no longer
valid for the regimes of high curvatures, and that quantum gravity corrections can be effectively accounted by the addition of higher order curvature terms
in the Lagrangian density for gravity. Significance of specific terms in these effective corrections will therefore determine the physical regime of cosmological evolution, 
finally leading to standard GR and $\Lambda$CDM model in the low curvature limit. We discuss the properties and establish the conditions for different
regimes in order to have a potentially viable model of cyclic cosmology. For simplicity we present the mathematical analysis in metric $f(R)$ gravity, but our
approach could as well be extended to other frameworks of modified gravity. This paper is organized as follows: in section 2. we present the idea of a cyclic
cosmology, its historical background and its recent developments. In section 3. we show the possibility of a simple modification of GR in the context of $f(R)$ gravity and 
its relationship to quantum gravity. In section 4. we carefully analyze the bouncing phase of the Universe with two different techniques, using perturbative calculation
and numerically. In section 5. we show how the inflationary phase can be incorporated into our model. In section 6. the present $\Lambda$CDM cosmology
is brought into a relationship with our model in the low curvature limit and the rip phase is analyzed as a necessary condition to avoid entropy issues during contraction.
In section 7. the final phase of a cycle, the contraction phase, is analyzed both analytically and numerically. In section 8. we 
reconstruct a concrete example of $f(R)$ function leading to cyclic model and in section 9. we conclude our work
 
\section{Models of cyclic cosmology}
 Although the idea of an eternally existing Universe, which undergoes infinite cycles of creation and destruction, reaches far back in the history of 
philosophy and mythology (probably even before the idea of Universe which is finite in time), in the context of physical cosmology it was first proposed in the 1930's by Tolmann \cite{tol}, and
also discussed by Lemaitre under the name of the ``phoenix Universe'' \cite{lem}. In the early Tolmann's idea this oscillatory behavior 
was enabled by the geometry of the Universe - it was the property of the solution of the Friedmann equations for the case of a closed Universe.
These first models were singular - after the initial singularity, the Universe would undergo a decelerated expansion, and after reaching a
turnaround point it would again collapse to itself. Apart from the existence of singularities, there are also intrinsic difficulties associated with this contracting phase. The collapse of matter would lead to a state of high densities and pressures leading to an unstable behavior, even more if the 
existence of black holes is taken into account - they would grow so large that the equations would break or cause a premature bounce \cite{lifshitz, misner, dicke}. 
Even if one could somehow resolve this stability and black holes issues, the collapse of matter would enormously increase the entropy of
the Universe. It would therefore increase by each cycle, and connecting this increasing entropy with increasing of the maximum radius 
of the Universe, one would get that radii were gradually smaller in the past. Thus, this argument would again lead to the initial singularity \cite{markov}.
With time it also became obvious that ingredients of these first cyclic models are not favored by the observations - it is currently 
believed that the Universe is flat rather than closed, so that Friedmann equations, assuming ordinary matter, do not lead to cyclic solutions.
Therefore, cosmological bounce and turnaround would require some unknown special physical mechanisms. For this reasons work on the cyclic cosmology
was mostly abandoned for decades. This situation changed with the discovery of accelerated expansion of the Universe in 1998., which opened an era of
extensive research on modified gravity theories. In fact, cosmological solutions with cyclic properties are not rarely encountered in different
theoretical frameworks of modified gravity \cite{cai2,ivanov,salah, varun, singh, biljar,sari}, and there are also numerous works analyzing the construction
of bouncing cosmologies in modified gravity \cite{amani, lopez,stach,moriconi, visser, roshan,noj-odin2,cai3, gurovich, ruter}. However, we will here be interested only in the full and 
potentially viable cyclic models. In our opinion, in order that cyclic models can be considered as viable it needs to 
lead to empirically observed aspects of $\Lambda$CDM cosmology in the current epoch - after the beginning of the radiation era until today - and
also solve the above stated problems connected to the contraction phase.\\ 
\\
Important attempts to address these issues started around 2001. with Ekpyrotic model as an alternative to inflation \cite{khoury}. The idea that dark energy could actually be a phantom energy, i.e. characterized
by equation of state (EOS) parameter $w=p/ \rho <-1$, with $\rho$ dark energy density and $p$ dark energy pressure, also played a significant role
in cyclic models. If dark energy is characterized by $w<-1$ then its energy density becomes so dominant that it destroys every gravitationally bounded system
until it diverges in the Big Rip singularity \cite{doom} (for a recent discussion on the effects of quantum fields and second-order curvature correction in the gravitational Lagrangian
on future singularities see \cite{carlson}). Using this fact, together with some mechanism that could start a contraction before the Big Rip
singularity is reached, the Universe could be treated as essentially empty at contraction, and then also black holes can disappear due to the failing of the Hawking
area theorem \cite{davies}. This assumption - together with the modification of Friedmann equations due to the string-theory ``braneworld'' scenarios  - was 
used in \cite{brown} to propose a new cyclic cosmological model. Other recently proposed types of models also typically use string-theory
framework to achieve bounce and turnaround, as can be reviewed in \cite{lehners}.  On the other hand, the cyclic model presented in 
\cite{steinhardt} achieves viability by introducing a negative scalar field potential and a coupling of this postulated
field with radiation and matter densities - which are however also motivated by string theory. Recently, the original Tolmann's idea of cyclic cosmology
based on a closed Universe has been revisited in \cite{george} using a phase plane analysis, with the addition of decaying cosmological term, and where also
the coupling with a scalar field has been introduced. The reconstruction of different $f(R)$ functions leading to cyclic solutions, corresponding to some
of the previously mentioned models, was analysed in \cite{gomez}. An interesting discussion on possible influences of contracting phase 
on cosmic microwave background asymmetries can be found in \cite{piao1,piao2} .
\\ 
\\
In our model we do not work within the string theory or any other specific framework, but assume that quantum gravity effects can be modeled by
higher order curvature corrections to the standard GR Lagrangian density for gravity. In this picture ``dark energy`` can be considered simply as
a zeroth contribution of these corrections. There is, moreover, no need for introducing any new hypothetical ingredients - like scalar fields, phantom energy and 
extra dimensions. Since the spacial curvature plays no role in supporting the cyclical behavior in our model, we take the generally accepted case 
where the Universe is flat.
We also take into account that coefficients associated with curvature corrections 
will in general be dependent on the curvature regime. Each cycle starts from the cosmological bounce, which ends the contraction
phase of the previous cycle, and then follows the evolution which subsequently leads to a decrease of the Ricci curvature scalar, such that higher 
order corrections become insignificant. After the standard $\Lambda$CDM era cosmology, the variation of the coefficients leads to a fast 
growth of effective ''dark energy`` contribution, causing a non-singular rip of bounded systems and the transition to a contraction phase of the Universe.
Then, during the contraction of an essentially empty Universe, higher order curvature terms again become significant, supporting the evolution that enables the transition to a new bounce and beginning
of a new cycle. We will examine each of the mentioned regimes and determine the necessary conditions and relations such that the above described
model is possible.

\section{$f(R)$ modified gravity}
 The lack of a complete understanding and theoretical motivation - which makes $\Lambda$CDM model more an empirical fit than a complete model - signifies the need for 
finding a quantized theory of gravity. This need supports the interest for investigation of modified theories of gravity. As
have been said, one of the simplest possible modification of GR is the $f(R)$ gravity. 
This theory generalizes the Lagrangian density in the Einstein-Hilbert 
action \cite{landau,misner0}:
\begin{equation}
 S_{EH}=\frac{1}{2\kappa}\int  d^{4}x\sqrt{-g}R,
\end{equation}
to become
a general function of the Ricci curvature scalar, $R$ \cite{buc}:
\begin{equation}
 S=\frac{1}{2\kappa}\int  d^{4}x\sqrt{-g}f(R),
 \label{akcija}
 \end{equation}
where $\kappa=8 \pi G$ ($c=\hbar=1$), $G$ is the gravitational constant and $g$ is the determinant of the metric. This modification can give a simple insight on 
modified gravitational effects and can serve as a toy theory of quantum gravity. The main interest in $f(R)$ theories comes from the possibility of explaining the
accelerated expansion of the Universe without introducing dark energy.
In fact, one can show that the vacuum solution of the field equations
in $f(R)$, for the special case $R=const.$, are the same as in ordinary GR with the addition of a cosmological constant, which then naturally results 
in a Schwarzschild-de Sitter Universe and the
accelerated expansion of the Universe \cite{faraoni}.
Also, as pointed out in \cite{cembranos} $f(R)$ theories with minimal modification lead to
new degrees of freedom and these new states can potentially provide the main contribution to the non-Baryonic dark matter. 
The $f(R)$ theory includes the four-derivative terms in the metric and
does not violate general covariance; moreover the four-derivative gravity is known to be renormalizable \cite{tombolis, steel, salam, julve, motola, fedkin, eliza}.
The $f(R)$ should in our opinion be considered as a toy theory, since it would be naive to think that a
real new theory of gravity could be constructed without changing the notion of space-time. The problem of quantum gravity will, in our opinion, not be solved by some random guess of the appropriate $f(R)$ function. Instead, 
the general foundational questions should be reconsidered with the new notions of dynamics and space-time learned from quantum mechanics and general relativity. Taking the $f(R)$
formalism as an effective toy theory for modeling the quantum effects on high curvatures, we will not be concerned here with specific forms of $f(R)$ 
function. Our approach will be to change the
GR with minimal modification in action similar to \cite{cembranos}, where we assume that $f(R)$ is analytic around $R=R_{0}$ and can be expanded in 
Taylor series \cite{psa, stabile, faulkner, clif, berry}:
\begin{equation}
 f(R)=c_{0}+\frac{c_{1}}{1!}(R-R_{0})+\frac{c_{2}}{2!}(R-R_{0})^{2}+\frac{c_{3}}{3!}(R-R_{0})^{3}+ \mathcal{O}((R-R_{0})^{4})   ...
 \label{taylorf}
\end{equation}
the dimension of $f(R)$ should be the same as $R$, it then follows that the dimension of $c_{i}$ is $[R]^{1-i}$. We will use this expansion in order to find 
conditions on coefficients in different regimes, which makes our approach general and non-dependent on specific models. \\
\\
We also take into consideration that the given Taylor expansions -- corresponding to different $f(R)$ function -- will in general be dependent on 
a specific curvature regime. It seems plausible that quantum corrections will effectively manifest in a different form, depending on the properties 
of the system under consideration. Otherwise, if one $f(R)$ function could cover all the scales and regimes, it could in principle already be considered 
as a complete theory of quantum gravity, and not just an effective description - which we assume it is not the case. We take this into account making the coefficients of the Taylor expansion 
 mildly dependent on curvature $c(R)$ - such that their variation can be neglected in a specific curvature region. To make it more suitable for 
our cyclic model we express (\ref{taylorf}) in terms of the maximal Ricci curvature of the Universe, $R_{max}$, and take $R_{0}=0$. Moreover, we identify $c_{0}$, 
as an effective cosmological term (which is now no longer constant but mildly dependent on curvature),
$c_{0}= - 2 \Lambda$ . Then the expansion (\ref{taylorf}) reads: 
\begin{equation}
 f(R)= - 2\Lambda(R) + \sum_{i=1}^{\infty} c_{i}(R) (R/R_{max})^{i},
 \label{suma}
\end{equation}
where we also made the substitution $c_{i} \rightarrow c_{i} R_{max}^i$. We note that the functional scale dependence of the couplings defining the theory, $c_{i}(R)$,
is to be expected from an effective field theory, as for example   analyzed in asymptotically safe approach towards quantum gravity \cite{sigurna1, sigurna2}.
The scale of interest is here naturally given by the curvature scalar, whose value determines UV and IR regimes of the theory. Since the Ricci curvature scalar
is a function of time in the cosmological context, it follows that these couplings will be time dependent. We emphasize that terms $\sum_{i=1}^{\infty} c_{i}(R) (R/R_{max})^{i}$
can simply be redefined to a new single function of Ricci scalar, which enters into Lagrangian density as any other $f(R)$ function, and therefore variational
procedure is still well defined. From a mathematical perspective any $f(R)$ function with appropriate behavior in specific curvature regimes, satisfying conditions
imposed on the expansion coefficients - which will be determined in the following sections, can support our cyclic model. However, reconstruction of a single
$f(R)$ function, satisfying these conditions in specific regimes, is beyond the scope of our work. Moreover, in our opinion it is unrealistic to expect
that a single $f(R)$ function could properly model effects of quantum gravity at all scales of interest. Therefore, the approach that we use should be
considered as a more general expansion in the Ricci curvature with running couplings, rather than a single $f(R)$ function formalism.
\\
\\
Modified $f(R)$ gravity, and therefore also the expansion (\ref{suma}), is known to be free from the Ostrogradsky instability \cite{woodard}. 
Moreover, the theory based on the expansion (\ref{suma}) will be stable and ghost-free if the following conditions are satisfied \cite{dolgov,sawicki, faraoni}
\begin{equation}
\frac{df(R)}{dR}> 0,
\end{equation}
\begin{equation}
 \frac{d^2 f(R)}{dR^2}\geq 0 .
\end{equation}

In order to obtain the field equations we will use the standard metric formalism. Beginning with the action (\ref{akcija}) adding the matter 
term, $S_{m}(g_{\mu\nu}, \phi)$, and varying
it with respect to the metric we get the modified field equations:
\begin{equation}
 f'(R)R_{\mu \nu} - \frac{1}{2}f(R)g_{\mu \nu} - (\nabla_{\mu} \nabla_{\nu} - g_{\mu\nu} \square)f'(R)=\kappa T_{\mu\nu},
 \label{fields}
\end{equation}
as usual the stress-energy tensor is:
\begin{equation}
 T_{\mu\nu}=\frac{-2}{\sqrt{-g}}\frac{\delta S_{m}}{\delta g^{\mu\nu}},
 \label{stress}
\end{equation}
where the prime denotes differentiation with respect to the argument, $\nabla_{\mu}$ is the covariant derivative, and $\square$ is $\nabla_{\mu}\nabla^{\mu}$.
The field equations (\ref{fields}) will be used in order to solve the modified Friedmann equations in different regimes.\\
\\
The structure of the field equations in $f(R)$ makes it possible to avoid singularities which are inevitable in standard GR. The necessary ingredient
of the singularity theorems is the requirement that the Strong Energy Condition (SEC) is satisfied \cite{elise}. Namely, if $\xi^{\mu}$ is a unit timelike four-vector
than SEC reads:
\begin{equation}
 \xi^{\mu}\xi^{\nu}T_{\mu \nu} \geq -\frac{1}{2}T,
\end{equation}
where $T$ is the trace of the stress-energy tensor defined in (\ref{stress}). Since all ordinary matter satisfies SEC, it appears that realistic cosmologies 
based on the standard GR must lead to singularities. 
However, in $f(R)$ modified gravity one can require that SEC for the matter stress-energy tensor is satisfied, while $R_{\mu\nu}\xi^{\mu}\xi^{\nu} < 0$.
If one defines the effective stress-energy tensor, $T_{\mu\nu}^{eff}$, which takes into account the effect of non-Hilbert terms in the action, then the equation 
(\ref{fields}) can be rewritten as:
\begin{equation}
 R_{\mu\nu}- \frac{1}{2}g_{\mu\nu}R=\kappa (T_{\mu\nu}+T_{\mu\nu}^{eff}).
\label{effective}
\end{equation}
It can be required that the total effective SEC can be violated:
\begin{equation}
 (T_{\mu\nu}+T_{\mu\nu}^{eff})\xi^{\mu}\xi^{\nu}<0,
\end{equation}
while matter components, $T_{\mu\nu}$, can still satisfy SEC. Under these conditions it is possible to avoid singularities in modified gravity, which is an essential
component of our cyclic model. \\
\\
We will take the assumption that the Universe is homogeneous and isotropic (the cosmological principle), which then is described by the FLRW line element in spherical
coordinates:
\begin{equation}
 ds^{2}=-dt^{2}+a(t)^{2}\bigg[  \frac{dr^{2}}{1-kr^{2}}+ r^{2}(d\theta^{2} +  \sin^{2} \theta d\phi^{2})  \bigg],
 \label{metrika}
\end{equation}
where $a(t)$ is the scale factor, $k=\pm1$ describes the spatial curvature, where $k=+1$ describing positive spatial curvature, $k=-1$ negative curvature and $k=0$ corresponds to 
local flat space. Current observations seem to favor the flat Universe, so in this work we will consider the case $k=0$ \cite{planco}. 
The matter in this model of the Universe will be described as
a perfect-fluid with the energy-momentum tensor:
\begin{equation}
 T_{\mu \nu}=(\rho + p)u_{\mu}u_{\nu} + pg_{\mu \nu},
 \label{idealni}
\end{equation}
where $\rho$ is the density, $p$ is the pressure, $u_{\mu}$ is the four-velocity and satisfies $u_{\mu}u^{\mu}=-1$.
The energy momentum conservation:
\begin{equation}
 \nabla_{\mu}T^{\mu \nu}=0,
\end{equation}
yields the equation:
\begin{equation}
 \dot{\rho}+3H(t)(\rho + p)=0,
 \label{conservation}
\end{equation}
where the dot is the time derivative and $H(t)=\frac{\dot{a}}{a}$ is the Hubble parameter.

\section{Bouncing phase}

 According to the framework of our previously described cosmological model, we assume that each cycle of the 
Universe begins from the contraction phase of the previous cycle, so that curvature, densities and all other 
physical quantities remain finite and well defined. Following already established terminology we will call this transition - from contraction in 
the previous cycle to expansion 
in the new cycle - bouncing phase of the Universe. This directly leads to some simple conditions for the Hubble parameter which need to be satisfied by 
any bouncing, and therefore
also the cyclical, cosmological model. If the scale factor of the Universe has a minimum at some moment $t_{0}$ (measured from some time, arbitrarily chosen, as an 
origin in infinite history of the Universe)
then, taking $d>0$ to be the time parameter, it follows i) $H(t_0 -d)<0$, ii) $H(t_{0} +d)>0$, iii) $H(t_{0})=0$ 
if $\lvert d - t_{0} \rvert < \lvert t_{max} - t_{0}\rvert$, where $t_{max}$ is the time 
where the scale factor reaches its maximal value. For simplicity, we can focus our attention on a special case which satisfies all
three conditions and assumes the symmetry of the scale factor around $t_{0}$: $H(t_0 -d)= - H(t_{0} +d)$.
In accord with the limit of cosmology based on the General relativity, our second physical requirement is that the Ricci curvature reaches its maximal value during the bouncing phase of the cycle,
at $t=t_{0}$. It can be shown in general that if the curvature scalar reaches its maximal and finite value at a certain point in time, FLRW geometry leads 
to the solutions which satisfy bouncing conditions ((i) - (iii)) on some interval around this point. If the Ricci scalar has a maximum at $t_{0}$, then for sufficiently small intervals of time around it,
$\lvert t - t_{0} \rvert \ll 1$, it can be described by first terms in the Taylor expansion:
\begin{multline}
6\dot{H} + 12H^2 = R_{max} + R_{2}(t - t_{0})^2 +R_{3}(t - t_{0})^3 + R_{4}(t - t_{0})^4 + \mathcal{O}((t-t_{0})^{5})...,
\label{vesela}
\end{multline}
where $R_{max} >0$, $R_{2}<0$ and $R_{max}$ is the curvature at the minimum of the scale factor, which happens at time $t_{0}$.
If we also assume that $R(t)$ is even around $t_{0}$, choosing $R_{3}=0$, one can also easily obtain the following conditions: 
\begin{itemize} 
 \item  $\dot{H}(t_{0} +d) = \dot{H}(t_{0}-d)$,
 \item  $\dot{H}(t) >0$ in the bouncing region,
 \item   $6\dot{H}(t_{0})=R_{max}$,
 \item  $\ddot{H}(t_{0})=0$,
 \item  $\ddot{R}(t_{0})=6\dddot{H}(t_{0}) + 24\dot{H}^{2}(t_{0})$. 
\end{itemize}
The equation $6\dot{H} + 12H^2 = R(t_{0})=R_{max}$ has the analytical solution:
\begin{equation}
H(t)=\frac{\sqrt{R_{max}}}{2 \sqrt{3}} \tanh \Big( \frac{\sqrt{3 R_{max}} t - C}{3} \Big),
\label{hiperbolni}
\end{equation}
with integration constant $C$. The complete equation (\ref{vesela}) can therefore be solved
by a perturbation procedure. It can be easily checked that this solution, consisting of (\ref{hiperbolni}) and small higher order corrections,
satisfies the bouncing conditions ((i) - (iii)) on a sufficiently small interval around $t_{0}$, and this also determines the value of the integration 
constant. Requirement of a small time interval leads to the physically plausible demand that bouncing phase needs to be small when compared to the period
of a complete cycle of the Universe.  We see that a bouncing solution in general appears as a rather natural feature of the FLRW geometry. Of course, this is only a geometrical consideration and the
real physical question is if the field equations for gravity admit solutions which lead to a Ricci scalar that satisfies Eq. (\ref{vesela}). In General relativity, which leads to an initial singularity in the curvature,
this condition is obviously not fulfilled. This shows that the paradigm of the beginning of the Universe (i.e. non-cyclicity of the Universe) in standard cosmology is entirely based 
on the specific mathematical form of the field equations, and not on the features of FLRW metric based on observations. This is important, since the field equations based on the same set of physical assumptions as the standard General relativity can in principle have
different form (as explored in modify gravity theories), as long as they are consistent, and lead to the same Newtonian and low-curvature limit. As we will show, even the introduction of very simple higher-order curvature terms
in the action integral for gravity can lead to field equations that satisfy (\ref{vesela}).
\\
\\
After these completely general remarks about the bounce in FLRW geometry, we turn to the proposed concrete model of cyclical Universe 
in general $f(R)$ gravity. In this model the bouncing phase comes after the rip phase, and is induced by an increase of the curvature which effectively activates higher terms in the expansion 
of $f(R)$ function. It is followed by the inflation phase (or possibly some other phase determined by the lower order in $f(R)$ expansion) during which curvature continues to
diminish, and which subsequently leads to the low-curvature phase of the standard cosmology, $f(R) \approx -2 \Lambda + R$. Therefore, in the bouncing phase one 
naturally needs to take into account the expansion of $f(R)$ function to the highest order, $N$, with respect to other phases. As we have already discussed, we take the bouncing phase to be
significantly shorter than the total time of one cycle of the Universe. Therefore, the coefficients $c_{n}$ can be taken to be constant on this small interval of time. 
Then starting from equations (\ref{suma}), (\ref{fields}), (\ref{metrika}), (\ref{idealni}) we obtain the modified Friedmann equation in the bouncing region:
\begin{multline}
3H(t)^{2} \sum_{n=0}^{N} c_{n} n R(t)^{n-1} = \rho(t)_{mat} + \rho(t)_{rad} \\ 
+ \frac{1}{2}[R(t) \sum_{n=0}^{N}c_{n} n R(t)^{n-1} 
- \sum_{n=0}^{N} c_{n}R(t)^{n}] \\
 - 3H(t) \sum_{n=0}^{N}c_{n}(n-1) n R^{n-2} \dot{R}(t),
\label{glavna}
\end{multline}
where $H(t)$ needs to be a solution of equation (\ref{vesela}) and $R(t)$ is given by the expansion given in the RHS of the same equation. The problem of constructing a bouncing phase
in general $f(R)$ gravity after its expansion in curvature terms, is therefore reduced to the problem of finding some set of coefficients $c_{n}$ which are consistent with (\ref{glavna}) and (\ref{vesela}) on some interval $d$ around $t_{0}$.  
Energy densities are given as a functions of the scale factor by the conservation of the stress-energy tensor, (\ref{conservation}), and can be expressed as a function of the Hubble parameter, according to $a(t) \sim e^{\int H(t) dt}$. Therefore any $f(R)$ theory 
whose factors of high-curvature expansion around $t_{0}$ satisfy (\ref{glavna}) and (\ref{vesela}) leads to a cosmological bounce, which is the necessary component of cyclical cosmology.
\\
\\
For simplicity we will consider this $f(R)$ expansion only to the third order in $R$. Higher orders introduce more parameters and therefore make the physical goals of the cyclical model, such as a non-singular and well behaved bounce,
in principle more easily obtainable - although leading to more complicated equations. On the other hand, our approach here will be
to construct a viable cosmological cyclical model in general $f(R)$ gravity in the simplest possible mathematical framework suitable for analytical treatment, with the smallest amount
of assumptions and free parameters as possible. If necessary all of the expansions taken here can be easily extended. 
We have discussed why the viable cyclical cosmological model
should have a rip phase (without a singularity), which brings perfect fluid densities to negligible levels and happens before the
contracting phase. 
It will also be assumed that we can effectively treat the Universe as empty at the end of the previous and the beginning of the next cycle - i.e. during the bouncing phase. 
Strictly speaking, at the beginning of the expanding phase (at the minimum of 
the scale factor) values of densities need to be set in accordance to empirical findings \cite{planco} according to the later evolution of densities in the FLRW model.
If we do not want to introduce special mechanisms of matter and radiation creation after the bouncing phase, one would expect that these values are reached by the end of the
contracting phase of the previous cycle, in which values of densities rise as the scale factor 
gets smaller. Therefore, not only that densities are actually not equal to zero at the beginning of the expanding phase, but they
are in fact at their maximum there. Nevertheless, we will for simplicity assume that in the bouncing region higher order curvature terms
dominate over densities which, in turn, can be neglected. This leads to the condition:
\begin{equation}
 \rho_{mat}^{max} + \rho_{rad}^{max} \ll \frac{1}{2}(f_{R}(R_{max}) - f(R_{max})) - 3H(t_{0}) \frac{df_{R}(R)}{dt}|_{R=R_{max}},
\end{equation}
where $\rho_{mat}^{max}$ and $\rho_{rad}^{max}$ are maximal values of matter and radiation densities. 
Since the minimum of the scale factor, $a_{min}$, is a free parameter of the model, 
we can in principle always choose it high enough to satisfy this condition according to $ \rho_{max}=\rho_{today} a_{min}^{-3(1+w)}$ as long as it stays smaller
than scales corresponding to the highest observed red-shift factor. 
\begin{figure}[t]
  \centering
  \includegraphics[scale=0.26]{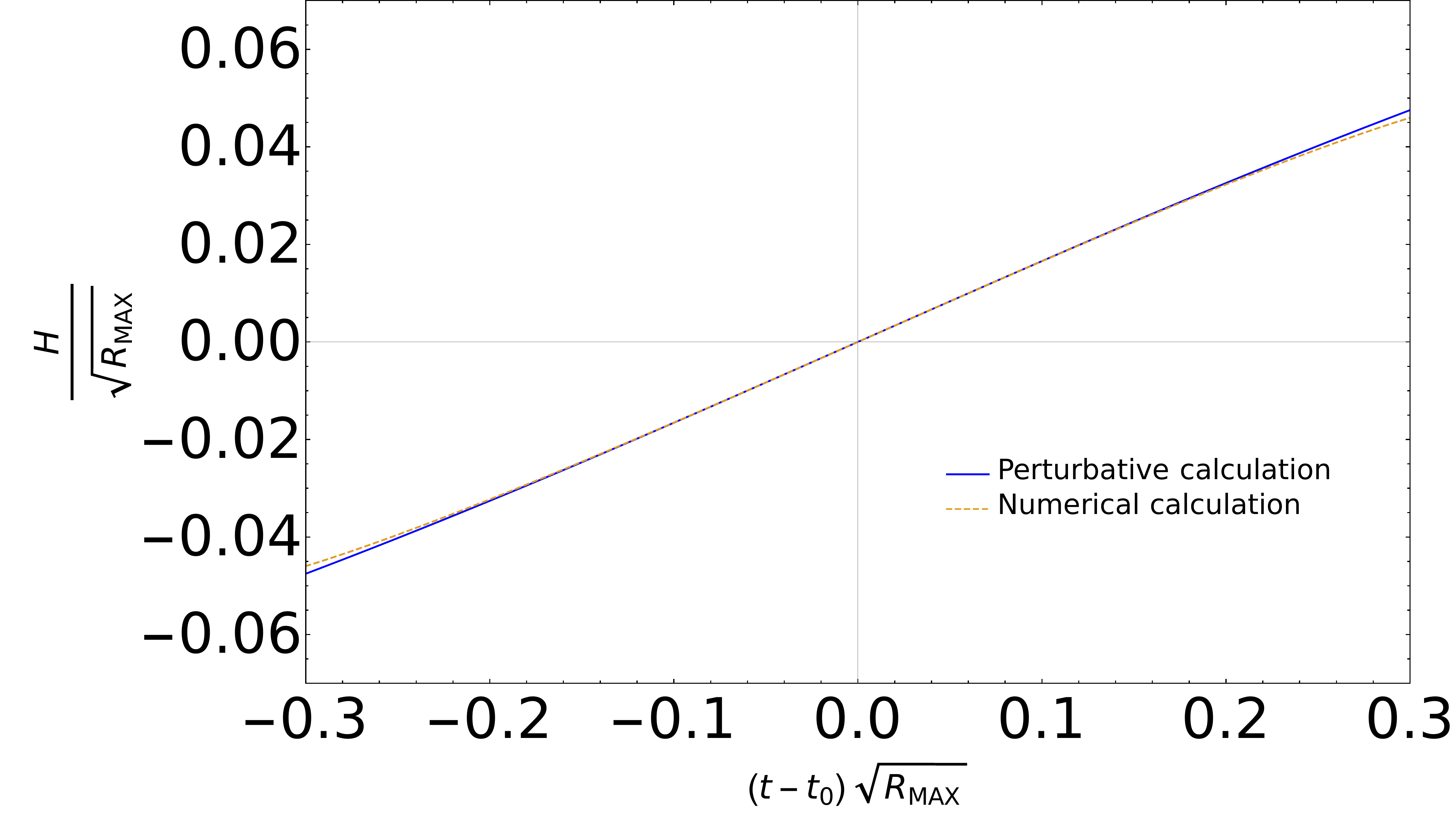}
  \caption{Time evolution of the Hubble parameter, $H(t)$, in the bouncing phase with parameters $\Lambda=0.0005$, $R_{2}=-2$.
  The perturbative solution (full line) is obtained from (\ref{vesela}) and (\ref{glavna}), using (\ref{par1})-(\ref{par3}).
  The numerical result (dashed line) is obtained by solving (\ref{numbounce}).}
  \label{hubbleslika}
  \end{figure}

\begin{figure}[h]
  \centering
  \includegraphics[scale=0.26]{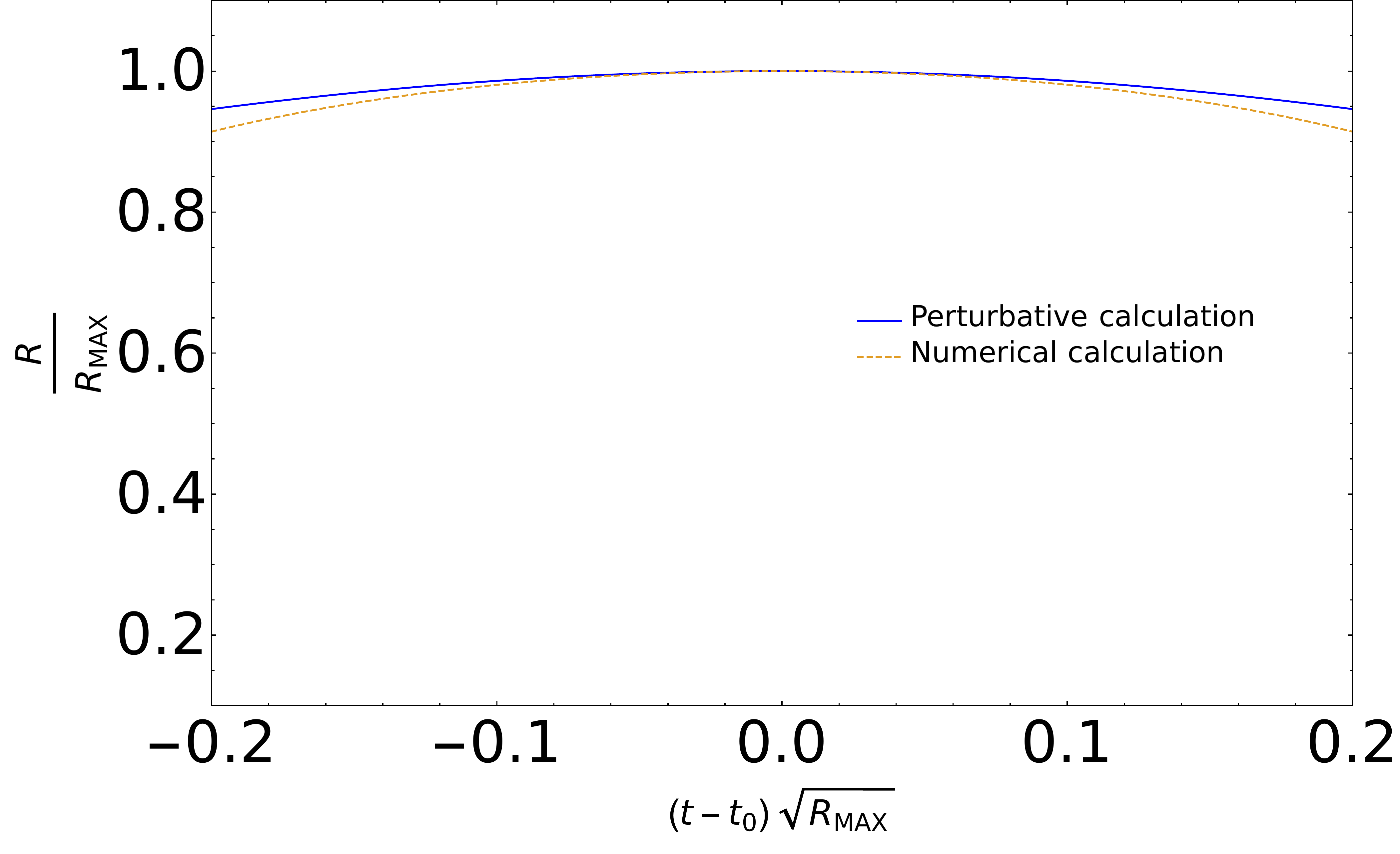}
  \caption{Time evolution of the Ricci curvature, $R(t)$, in the bouncing phase with parameters $\Lambda=0.0005$, $R_{2}=-2$.
  The perturbative solution (full line) is obtained from (\ref{vesela}) and (\ref{glavna}), using (\ref{par1})-(\ref{par3}).
  The numerical result (dashed line) is obtained by solving (\ref{numbounce}).}
  \label{riccislika}
  \end{figure}
  
Thus, expanding the $f(R)$  to the third other in $R$, and solving the equations  (\ref{vesela}) and (\ref{glavna}) to the
fourth order in time for the coefficients of $f(R)$ expansion, we obtain the following relations:
\begin{equation}
R_{4}=-\frac{R_{2}(3\Lambda -1 + 45 \Lambda R_{2}-24 R_{2}+324 \Lambda R_{2}^{2})}{18(3\Lambda -1)},
\label{par1}
\end{equation}
\begin{equation}
c_{2}=\frac{3 \Lambda +36 \Lambda R_{2}-2}{1-12R_{2}},
\label{par2}
\end{equation}
\begin{equation}
 c_{3}=\frac{-2 \Lambda-12 \Lambda R_{2}+1}{1-12 R_{2}},
 \label{par3}
\end{equation}
where for simplicity (and easier comparison with the standard GR) we took $c_{1}=1$. We present solutions for time evolution of
$H(t)$ and $R(t)$ corresponding to parameters (\ref{par1})-(\ref{par3}) in Fig. \ref{hubbleslika} and Fig. \ref{riccislika}.
Also, in the same figures we compare this analytical perturbative solutions with the numerical results in the bouncing phase to get more accurate and reliable results.
Numerical solutions are obtained by solving the full modified Friedmann equation:
\begin{equation}
 3H(t)^{2}\frac{df}{dR}-\frac{1}{2}\Big( R(t) \frac{df}{dR} -f(R) \Big) + 3 H(t) \frac{d^{2}f}{dR^{2}}\dot{R}(t)=0,
 \label{numbounce}
\end{equation}
taking $f(R)$ as in (\ref{suma}), expanded to the third order.
The field equation is a second order differential equation in $H(t)$, in order to solve it we need two initial conditions.
From the bounce definition at $t=t_{0}$ the Hubble parameter should be $H(t_{0})=0$. 
We note that in a small time region around the maximum the Ricci scalar appears to be symmetric. This comes from the fact
that for small intervals the second order term in Taylor expansion of $R(t)$ will be dominant, but for larger intervals
the Ricci scalar is not necessary symmetric in our model.
It can be seen that the numerical results are matching the perturbative method. While perturbative and numerical solutions are in an exellent agreement for 
the Hubble parameter, we see that for large $(t-t_{0})\sqrt{R_{Max}}$ the respective Ricci scalars start to slightly differ. This comes from the fact that for larger time intervals
higher order terms become significant, and  this error accumulates in the Ricci scalar which we compute from the $H(t)$ obtained by the perturbative calculation.\\
The transition through a cosmological bounce
signifies the beginning of a new cycle of the Universe, with the evolution that will be subsequently described by lower curvature regimes.

  \section{Early expansion phase}
  In principle, the bouncing phase - characterized by the significance of the highest order curvature corrections, can be 
followed by some phase of lower order Lagrangian corrections. In order that this phase matches the bouncing phase and the
standard radiation era it follows that $\dot{R}(t)<0$ and $\dot{a}(t)>0$ during this phase. Any set of parameters in the $f(R)$ expansion leading to these
conditions and solving the modified Friedmann equation is therefore consistent with the description of this phase. More specific,
although inflationary expansion may be conceptually unnecessary in the framework of cyclic Universe it can easily be embedded in our model. 
It would then just be a result of modifying gravity and not supported with any kind of special field. In fact, historically it was 
the Starobinsky's proposal for the inflation model that brought attention to the $f(R)$ modified gravity \cite{starob}. We will therefore not
study this phase in detail here, but only give a simple demonstration of this, taking a second order curvature correction
and assuming that a variation of $c_{i}$ can be neglected in this phase:
\begin{equation}
f(R)=- 2\Lambda + (R/R_{max}) + c_{2} (R/R_{max})^{2}. 
\label{early}
\end{equation}
Assuming that during this early expansion phase there is an era where $R(t)$ changes very slowly we can take 
$R(t)/R_{max} \equiv R_{c} \approx const.$ Then if $c_{2}R^{2} + \Lambda \gg \rho_{mat}(t) + \rho_{rad}(t)$ it follows
\begin{equation}
H^{2}=\frac{c_{2} R_{c}^{2} + 2\Lambda}{6(1+2c_{2}R_{c})}, 
\end{equation}
leading therefore to the inflationary growth of the scale factor after an initial time, $t_{in}$, given by
\begin{equation}
a(t)=a(t_{in})e^{\sqrt{\frac{c_{2} R_{c}^{2} + 2\Lambda}{6(1+2c_{2}R_{c})}}(t-t_{in})} .
\end{equation}

  \section{Phase of standard cosmology and the rip phase}
\begin{figure}[t]
  \centering
  \includegraphics[scale=0.23]{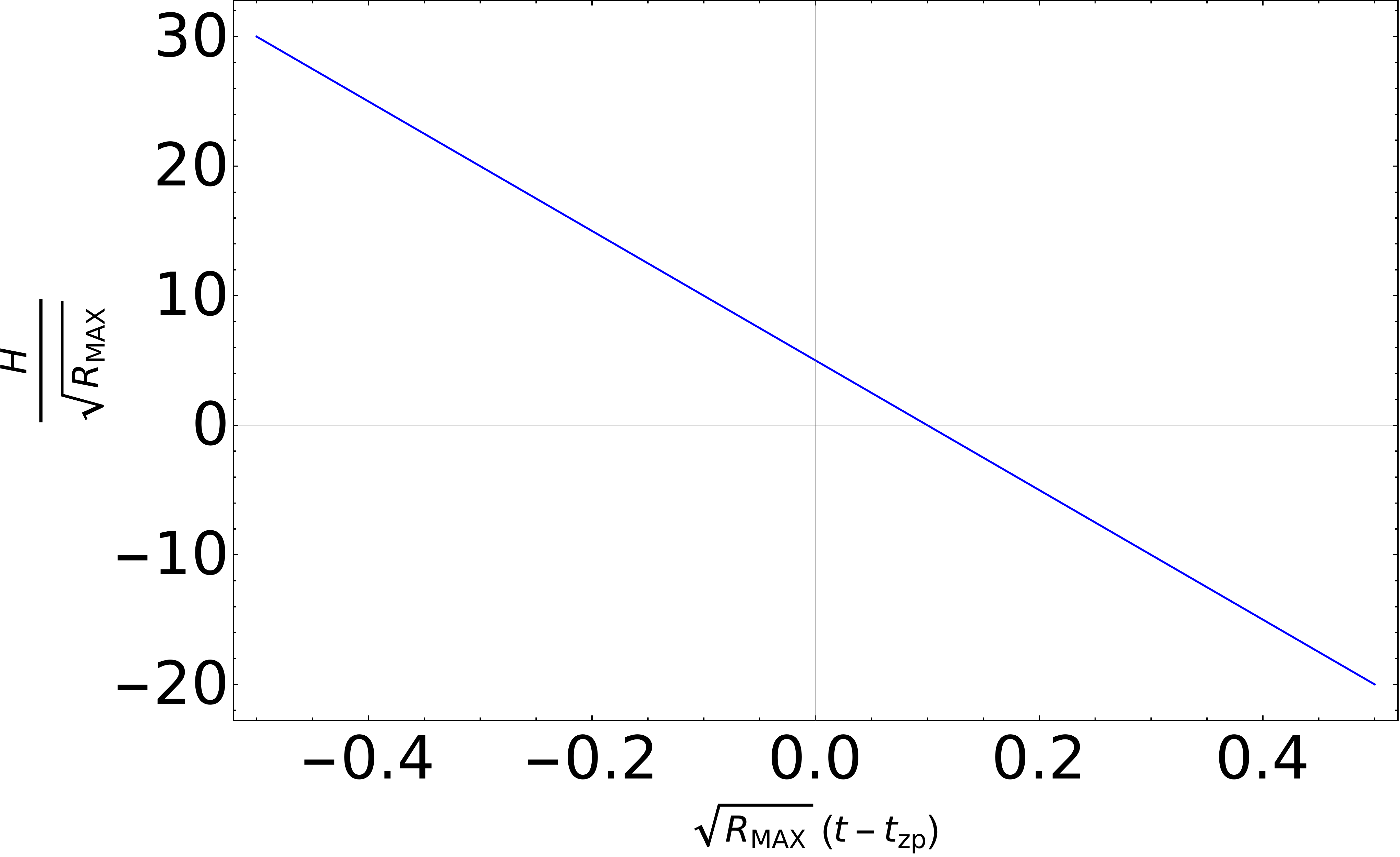}
  \caption{Time evolution of the Hubble parameter, $H(t)$, near the maximum of the scale factor in the rip phase, obtained as a solution of (\ref{ripanje}) with $b=5$.}
  \label{Hrip}
  \end{figure}

\begin{figure}[t]
  \centering
  \includegraphics[scale=0.24]{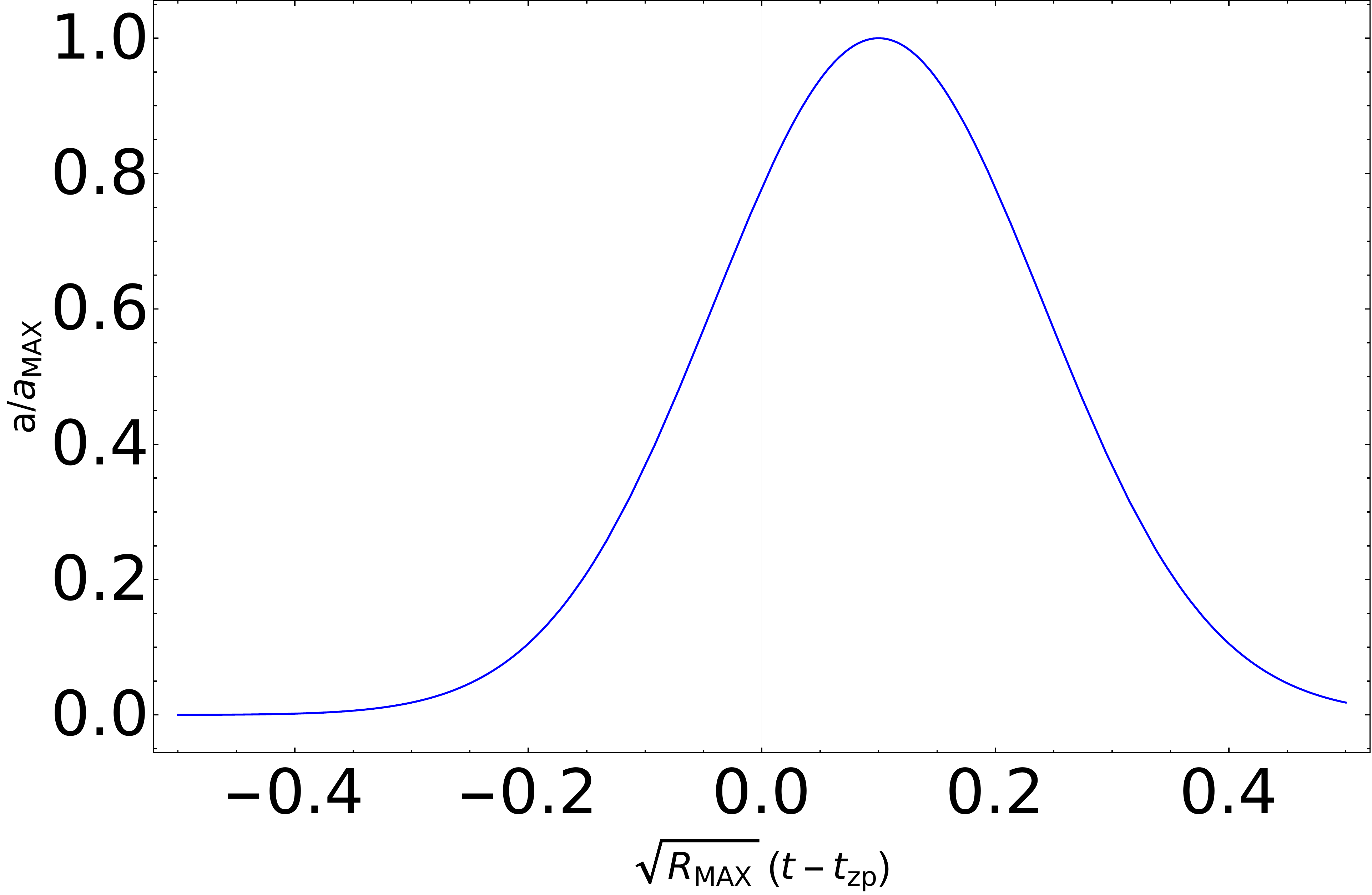}
  \caption{Time evolution of the scale factor, $a(t)$, near its maximum in the rip phase, with $b=5$.}
  \label{arip}
  \end{figure}

\begin{figure}[t]
  \centering
  \includegraphics[scale=0.22]{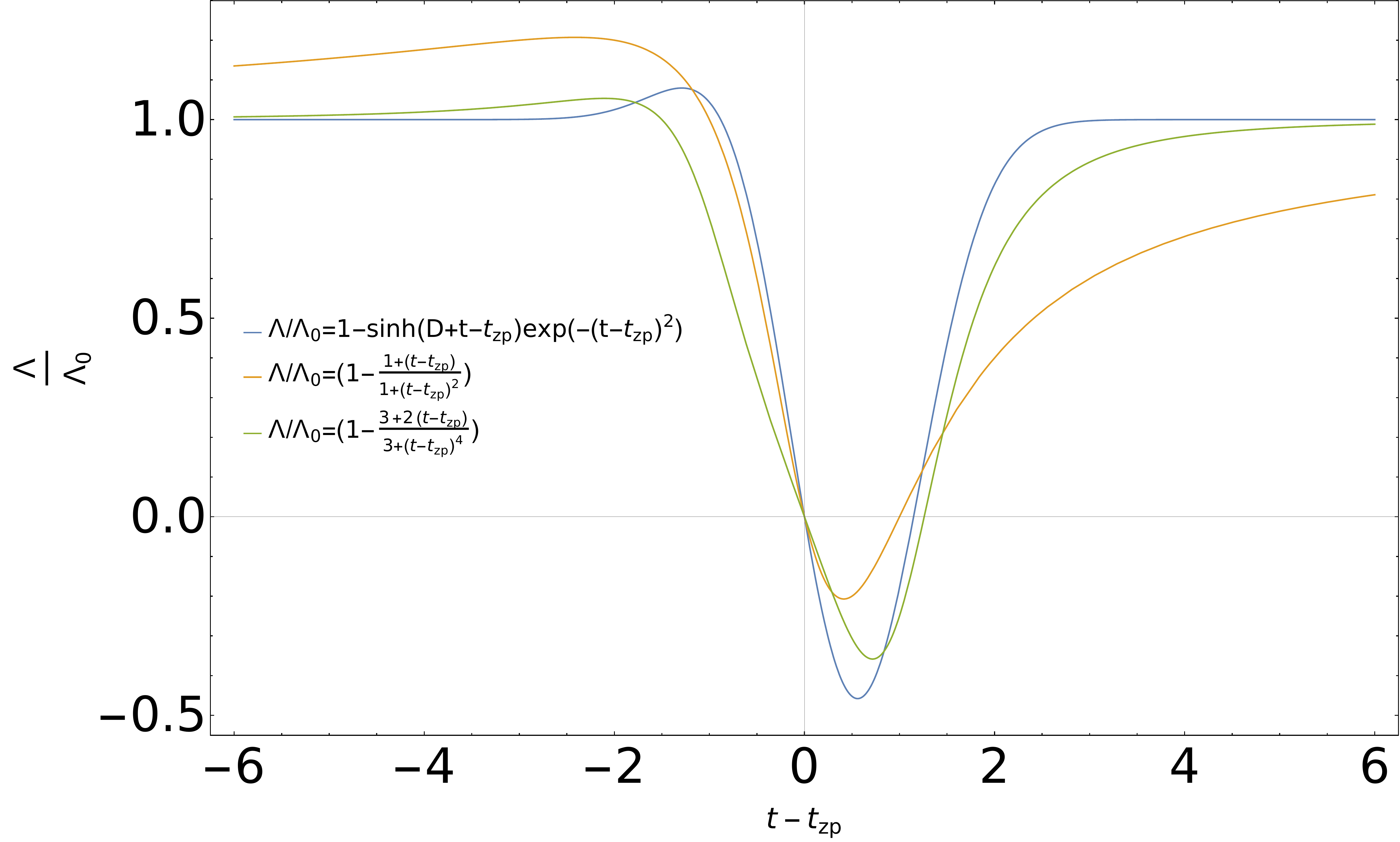}
  \caption{Example of the family of functions $\Lambda(t)$ which satisfy conditions  i) $\lim_{t \to \infty} \Lambda(t)= \Lambda_{0}=const.$, ii) $\lim_{t \to 0} \Lambda(t)=\Lambda_{0}$, iii)
$\dot{\Lambda}(t)>0$ for $t_{today}<t<t_{rip}$, iv) $\Lambda(t) \approx c(t-t_{zp})$ for $|t-t_{rip}| \ll |t_{rip}- t_{today}|$, with $D=\ln (1+\sqrt{2})$. }
  \label{Lrip}
  \end{figure}

 When the value of the Ricci curvature scalar of the Universe becomes small enough, $R/R_{max} \ll1$, all higher terms in $f(R)$
expansion are negligible and the evolution is governed by the $f(R)= - 2 \Lambda + R$ function. In order to have the
observed evolution of the Universe - with a radiation dominant era, matter dominant era and structure growth, as well as
the late accelerated expansion - this regime of a small curvature must last for a sufficiently long time. Therefore, on this time interval the coefficients $c_{i}$ 
should not be considered as constants - as was the case with the earlier phases, but rather as slowly varying functions
of $R(t)$. In the following, we will assume for simplicity that this time dependence can be completely absorbed in the $\Lambda(t)$ function and ignore
time variation of other coefficients. However, considering the case where $t_{today} -t_{bounce} \ll t_{max} - t_{bounce}$, $\Lambda(t)$ can still be considered
as a constant on time scales from bounce till today. We remind the reader that $t_{max}$ is the time where the scale factor reaches its maximal value.  
In this time region standard cosmological results are obtained, with an approximately
constant zero order-term of $f(R)$ expansion playing the role of the standard cosmological term. After the era of radiation and matter
domination, this term starts to dominate and leads to the long period of accelerated expansion of the Universe. The matter, $\rho^{mat}_{today}(a/a_{today})^{-3}$, 
and radiation, $\rho^{rad}_{today}(a/a_{today})^{-4}$, content
become negligible and after long enough time the Universe can be treated as essentially empty, with its dynamics completely determined by the $\Lambda$ term. On 
time scales of this assumed long accelerated expansion, the time variation of $\Lambda$ term now becomes essential.
In order to more easily study this regime, Friedmann equations can now be rearranged to read:
\begin{equation}
H^{2}=\frac{\kappa}{3}(\rho + \rho_{eff}),
\label{rip1}
\end{equation}
\begin{equation}
\dot{H}+H^{2}=-\frac{\kappa}{6}\Big(\rho + \rho_{eff} + 3(p+p_{eff})\Big),
\label{rip2}
\end{equation}
  where $p=(1/3) \rho_{rad}$ and we have introduced the effective energy density, $\rho_{eff}$, and effective pressure, $p_{eff}$, which describe the
contribution of the non-Hilbert term, $\Lambda(R)$, in the action integral for gravity. These effective terms do not correspond to any real physical fluid, 
but just effectively 
model the influence of the generalization of the standard field equations. The obvious advantage of this notation is that this model of modified gravity can be
formally compared to other models, also the ones including quintessence, scalar fields etc. From the modified Friedmann equation (\ref{numbounce}) with 
$f(R)=-2 \Lambda(R) + R$ using the chain rule, $d\Lambda(t)/dR(t)=\dot{\Lambda}(t)/\dot{R}(t)$, and comparing it with equation (\ref{rip1}) and (\ref{rip2}) we determine that the effective terms are given by:
\begin{equation}
 \rho_{eff}= \frac{1}{\kappa \Big(1 - \frac{2\dot{\Lambda}(t)}{\dot{R}(t)}\Big)} \Big[ \Lambda(t)-R(t) \frac{\dot{\Lambda}(t)}{\dot{R}(t)}
+ 6\Big( \frac{\ddot{\Lambda}(t)}{\dot{R}(t)}-\frac{\dot{\Lambda}(t)\ddot{R}(t)}{\dot{R}^{2}(t)} \Big)H(t) + \kappa \rho \frac{2\dot{\Lambda}(t)}{\dot{R}(t)}\Big] ,
\label{efen}
\end{equation}
\begin{multline}
p_{eff}=-\frac{1}{3\kappa \Big(1-\frac{2\dot{\Lambda}(t)}{\dot{R}(t)}\Big)}\Big[ R(t) \Big(1- \frac{\dot{\Lambda}(t)}{\dot{R(t)}} - \frac{\Lambda(t)}{R(t)}\Big)
 \\ -6 \Big( \frac{\ddot{\Lambda}(t)}{\dot{R}(t)}-\frac{\dot{\Lambda}(t)\ddot{R}(t)}{\dot{R}^{2}(t)} \Big)H(t)  
- \kappa \Big(\rho - 3p\Big(1-\frac{2\dot{\Lambda}(t)}{\dot{R}(t)}\Big)\Big)\Big] .
\label{efpres}
\end{multline}
Since the Universe can be considered as nearly empty, in the remaining discussion we will take $\rho=0$, and the effective equation of state (EOS) can then 
be written as:
\begin{equation}
w_{eff}(t)=-\frac{1}{3}\frac{ R(t) \Big(1- \frac{\dot{\Lambda}(t)}{\dot{R(t)}} - \frac{\Lambda(t)}{R(t)}\Big)
-6 \Big( \frac{\ddot{\Lambda}(t)}{\dot{R}(t)}-\frac{\dot{\Lambda}(t)\ddot{R}(t)}{\dot{R}^{2}(t)} \Big)H(t) }{\Lambda(t)-R(t) \frac{\dot{\Lambda}(t)}{\dot{R}(t)}
+ 6\Big( \frac{\ddot{\Lambda}(t)}{\dot{R}(t)}-\frac{\dot{\Lambda}(t)\ddot{R}(t)}{\dot{R}^{2}(t)} \Big)H(t) } .
\label{w}
\end{equation}
For time intervals such that $t\ll (t_{max} - t_{bounce})$ one can take $\Lambda \approx const.$ and then $w_{eff} \approx -1$. For these
early times we therefore obtain the standard cosmological term with the usual EOS. 
\\
\\
In the cyclical model, the Universe needs to reach the maximal value of the scale factor at time $t_{max}$ and then
enter into the contracting period of the cycle, finally leading to the bouncing phase and beginning of a new cycle. If this contracting phase would immediately follow
after the phase of accelerated expansion with $\Lambda=const.$ it would lead to the collapse of galaxies and matter of the Universe, new phase transitions,
and therefore enormous increase of the temperature end entropy - which would then be progressively higher at the beginning of every new cycle. 
In order to prevent these problems we assume that prior to entering into the contraction phase, all structures of the Universe need to be thorn apart due to 
the contribution of modified terms
(\ref{efen}) and (\ref{efpres}). Moreover, also due to the long period of accelerated expansion, we treat the Universe as empty when entering into the contraction phase. 
Therefore we call this phase of the cycle - a ripping phase, with the crucial difference with respect 
to the previously proposed Big rip scenarios \cite{doom, novi, novi2} that the scale factor remains always finite. 
This means that the contribution of the effective energy density contained in the modified terms needs to increase with time, in order to become significantly large
to destroy all bounded systems in the Universe. 
Starting from (\ref{efen}), (\ref{efpres}) and (\ref{w}) modified Einstein's equation can be written as in (\ref{effective}). Using the conservation of the
energy-momentum tensor for matter, and $\nabla_{\mu} G^{\mu \nu}=0$, it also follows that $ \nabla_{\mu} T_{\mu \nu}^{eff}=0$, leading to the equation
for the evolution of the effective energy density component
\begin{equation}
\dot{\rho}_{eff}(t)+ \frac{\dot{a}(t)}{a(t)} \rho_{eff}(t)(1+w_{eff}(t)) =0
\end{equation}
On the small interval around some time instant, $t=t_{i}$, $w_{eff}$ can be considered as constant, and it is then easy to see that
the effective energy contribution will grow on that interval if $w_{eff}< -1$. 
Therefore, for the late times it follows that $w_{eff}(t)< -1$, so that the effective energy contribution be always growing -- in order to
destroy all bounded systems. Although this represents the condition for 
phantom energy EOS,
our framework differs from various phantom energy proposals both physically and mathematically. The physical difference comes from the fact that
 we do not assume the existence of any
substance that would have the properties of the phantom energy, but this effective behavior comes 
just as a result of generalizing the action for GR. The important mathematical difference lies in the fact that the structure of the field equations
is essentially different from the case when one would assume the existence of the time dependent phantom energy contribution
to the energy-momentum tensor of standard General relativity. In the later case the system would be described by the equation
\begin{equation}
 H^{2}=\kappa \Big(\rho + \rho_{\Lambda}^{today}(\frac{a}{a_{today}})^{-3(1+w_{\Lambda})}\Big),
\end{equation}
with $w_{\Lambda}=const.$ instead of our equations (\ref{rip1}), (\ref{efen}) and (\ref{efpres}).
\\
\\
Since $\rho +3p$ can be considered as a source of gravitational potential, one can make an estimate of the rip time of a gravitationally 
bounded system. If we have an orbit around a mass $M$ then the rip time is roughly determined by the condition: $(\frac{4 \pi }{3}r^{3})[\rho_{eff} + 3p_{eff}] \approx M$. 
Therefore the rip time, $t_{rip}$, for an astrophysical object determined by the typical values for $M$ and $r$ can be written as:
\begin{equation}
 \frac{4 \pi r^{3}}{3\kappa}\frac{2 \Lambda(t_{rip}) - R(t_{rip})+12 \frac{d^{2}\Lambda(t)}{dR^{2}}H \dot{R}(t)}{1 - \frac{2\dot{\Lambda}(t_{rip})}{\dot{R}(t_{rip})}} \approx M.
\label{rip}
\end{equation}
In accord with the previous discussion, we require that $t_{rip}$, which will depend on a specific form of the $\Lambda(R)$
function, satisfies the condition $t_{rip} \leqslant t_{max}$. 
In order to have the transition from the expanding, $H(t)>0$, to the contracting phase $H(t)<0$, the scale factor must reach its
maximum value, which leads to the conditions:
\begin{equation}
\frac{\Lambda(t_{max})}{\dot{\Lambda}(t_{max})}=\frac{R(t_{max})}{\dot{R}(t_{max})},
\label{extr.condition}
\end{equation}
\begin{equation}
\rho_{eff}(t_{max}) + 3p_{eff}(t_{max}) >0 .
\label{max.condition}
\end{equation}
It is clear that a viable model of cyclic cosmology can not be obtained by adding new fluid components with a fixed EOS to the $\Lambda$CDM ones. 
Effective EOS needs to be a time dependent variable, changing from $w_{eff}<-1$ at $t_{rip}$ to $w_{eff}>-1/3$ at $t_{max}$, as given by (\ref{max.condition}). 
There is a broad potential class
of $\Lambda(t)$ functions that can satisfy this requirement, together with (\ref{rip}), (\ref{extr.condition}), (\ref{max.condition}). For simplicity we will here focus on
functions which also have the following properties: i) $\lim_{t \to \infty} \Lambda(t)= \Lambda_{0}=const.$, ii) $\lim_{t \to 0} \Lambda(t)=\Lambda_{0}$, iii)
$\dot{\Lambda}(t)>0$ for $t_{today}<t<t_{rip}$, iv) $\Lambda(t) \approx c(t-t_{zp})$ for $|t-t_{rip}| \ll |t_{rip}- t_{today}|$, where $c$ is a real number. 
These functions therefore all have a zero point, $t_{zp}$, and sufficiently far away from it can be 
considered as constants. In this case the influence
of the time change in the $\Lambda$ term can be ignored for all other phases of a cycle of the Universe. Some examples of functions with these properties include:
$\Lambda_{0}\Big( 1- \frac{1+(t-t_{zp})}{1+(t-t_{zp})^{2}}
\Big) $,
$\Lambda_{0}(1 - \sinh \Big(\ln(1+\sqrt{2})+(t-t_{zp})\Big)e^{-(t-t_{zp})^2} $
... 
\\ \\
In order to solve the Friedmann equation we should choose a $\Lambda$ that satisfy the above mentioned requirements. Physically
interesting regions are those in the vicinity of $t_{zp}$ where we approximate $\Lambda(t) \approx c(t-t_{zp})$ and the asymptotical
region $t \rightarrow \infty$ where we take $\Lambda(t)$ as a constant, $\Lambda_{0}$. 
\\
Solving the Friedmann equation with $\Lambda(t)=\Lambda_{0}$ gives the well known result from General Relativity with empty space 
filled with an effective cosmological constant:
\begin{equation}
 H^{2}=\frac{\Lambda}{3},
\end{equation}
as it should be if we want a smooth transition from the phase of the standard late time cosmology.
\\
Now we want to find the solution near $t_{zp}$, where approximately $\Lambda(t) = c(t-t_{zp})$. By inserting $\Lambda(t)$ into
Friedmann equation we get:
\begin{equation}
 6\dot{R}H^{2}=c\Big( 6\frac{\ddot{R}}{\dot{R}}H + 6H^{2} - R + \dot{R}(t-t_{zp})\Big).
 \label{ripanje}
\end{equation}
It can be shown that $H(t)=A(t-t_{zp}) + b$ is a solution of (\ref{ripanje}) on a sufficiently small interval around $t_{zp}$, where higher orders 
of $(t-t_{zp})$ are negligible.
From the field equation we get the following conditions:
\begin{equation}
 A=-2b^{2},
\end{equation}
\begin{equation}
 c=-48 b^{3}.
\end{equation}
We depict 
 $H(t)$, $a(t)$ and various $\Lambda(t)$ in figures \ref{Hrip}, \ref{arip} and \ref{Lrip}.
  \\ \\
After reaching the maximal value of the scale factor, the Universe will contract following the same dynamics governed by
(\ref{rip1}), (\ref{rip2}) until the value of the Ricci curvature becomes high enough so that the term $(R/R_{max})^{2}$
is again significant.  

\clearpage

\section{Transition to a new bounce}
\begin{figure}[h]
  \centering
  \includegraphics[scale=0.24]{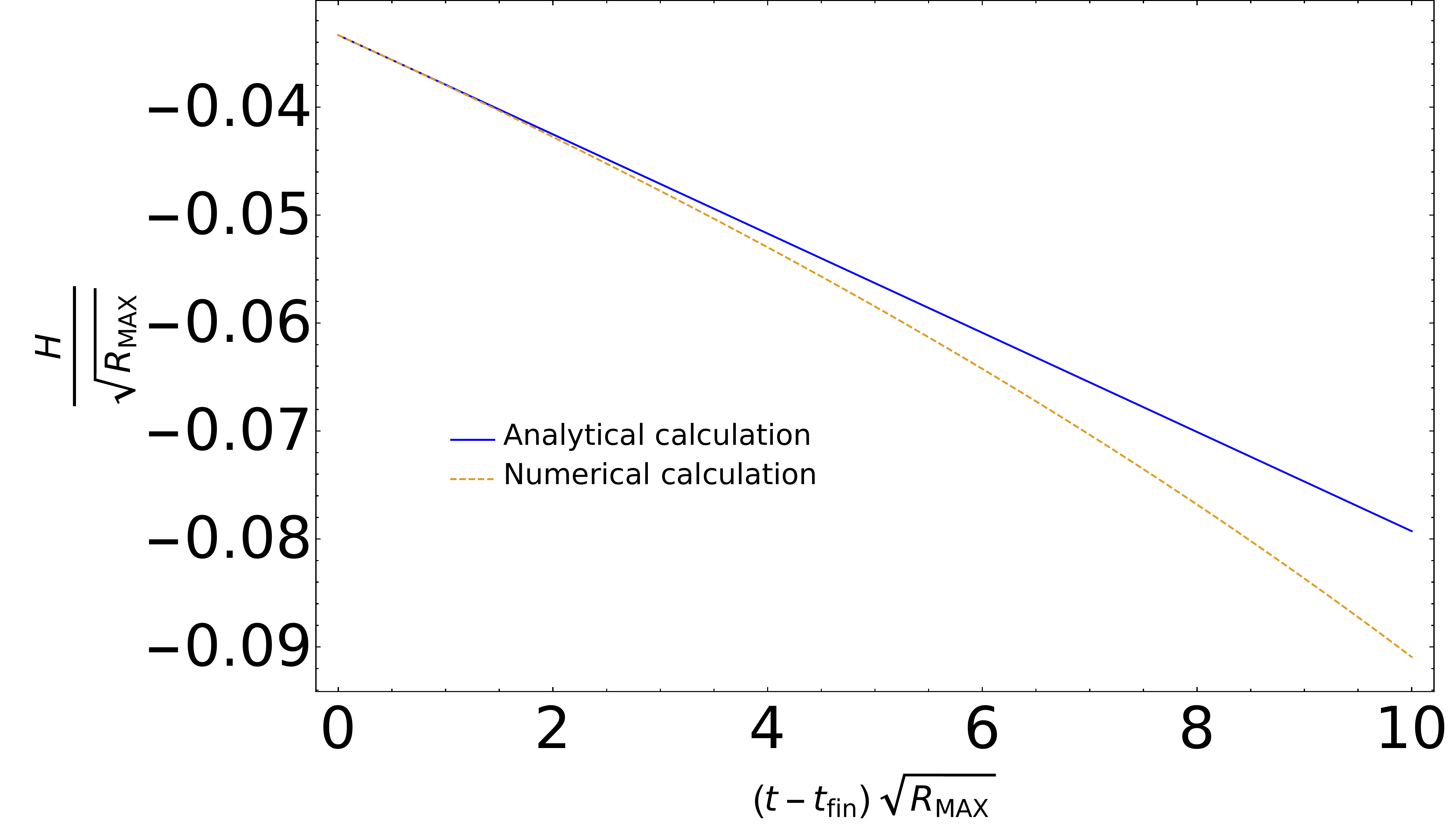}
  \caption{Time evolution of the Hubble parameter, $H(t)$, during the transition to a 
  cosmological bounce following (\ref{hubl-skupljanje}), with parameters $A_{1}=-0.005$, $H_{fin}=-0.03$ with
  $N=1$.
   The numerical result (dashed line) is obtained by solving (\ref{frid-skupljanje}) with $\Lambda=0.01$, $A_{1}=-0.005$, $H_{fin}=-0.03$, 
   using (\ref{par4}) and (\ref{par5}).}
  \label{hubbleslikaanti}
  \end{figure}

\begin{figure}[h]
  \centering
  \includegraphics[scale=0.24]{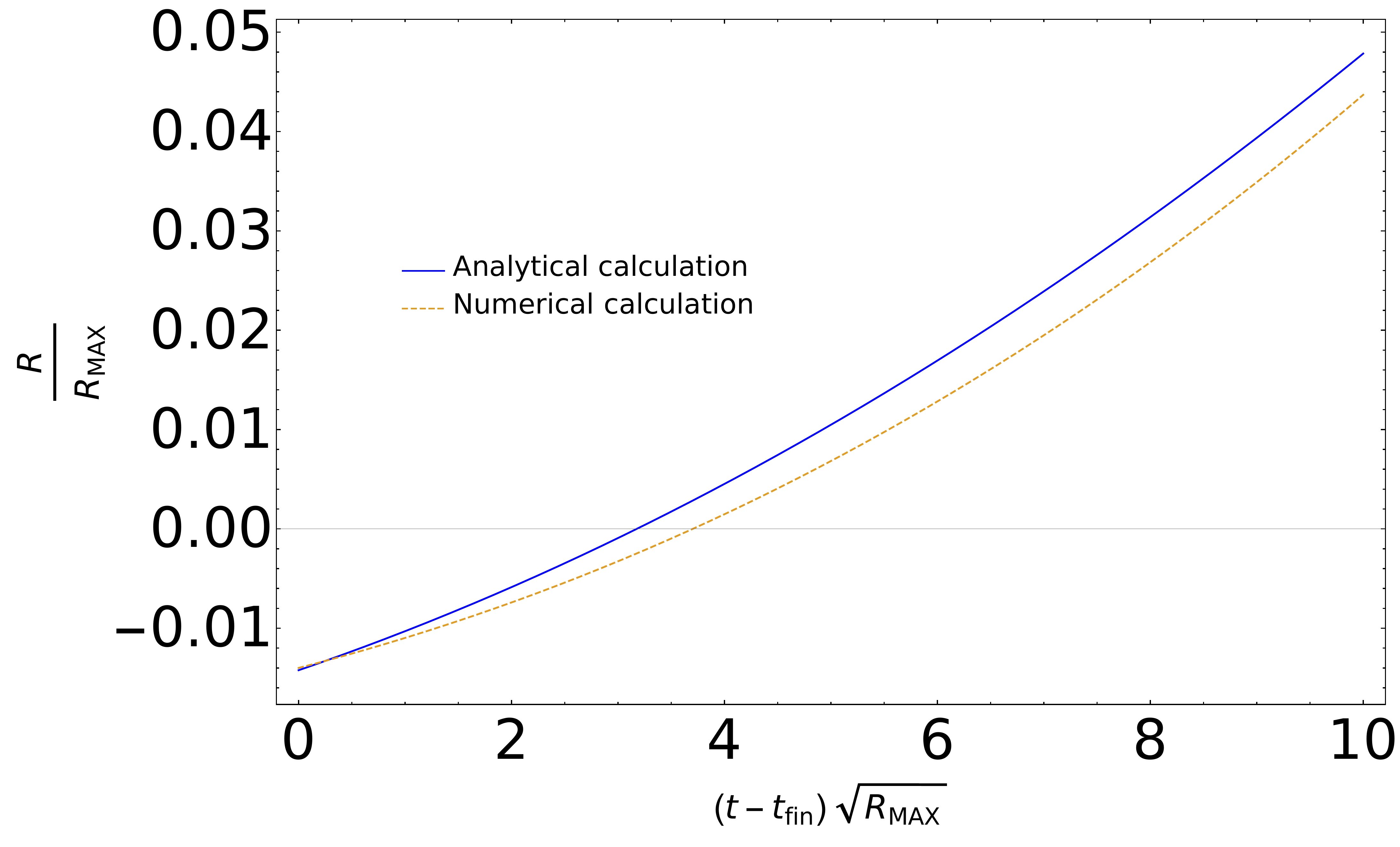}
  \caption{Time evolution of the Ricci curvature, $R(t)$, during the transition to a 
  cosmological bounce following (\ref{hubl-skupljanje}), with parameters $A_{1}=-0.005$, $H_{fin}=-0.03$ with
  $N=1$.
   The numerical result (dashed line) is obtained by solving (\ref{frid-skupljanje}) with $\Lambda=0.01$, $A_{1}=-0.005$, $H_{fin}=-0.03$, 
   using (\ref{par4}) and (\ref{par5}).}
  \label{riccislikaanti}
  \end{figure}

\begin{figure}[h]
  \centering
  \includegraphics[scale=0.24]{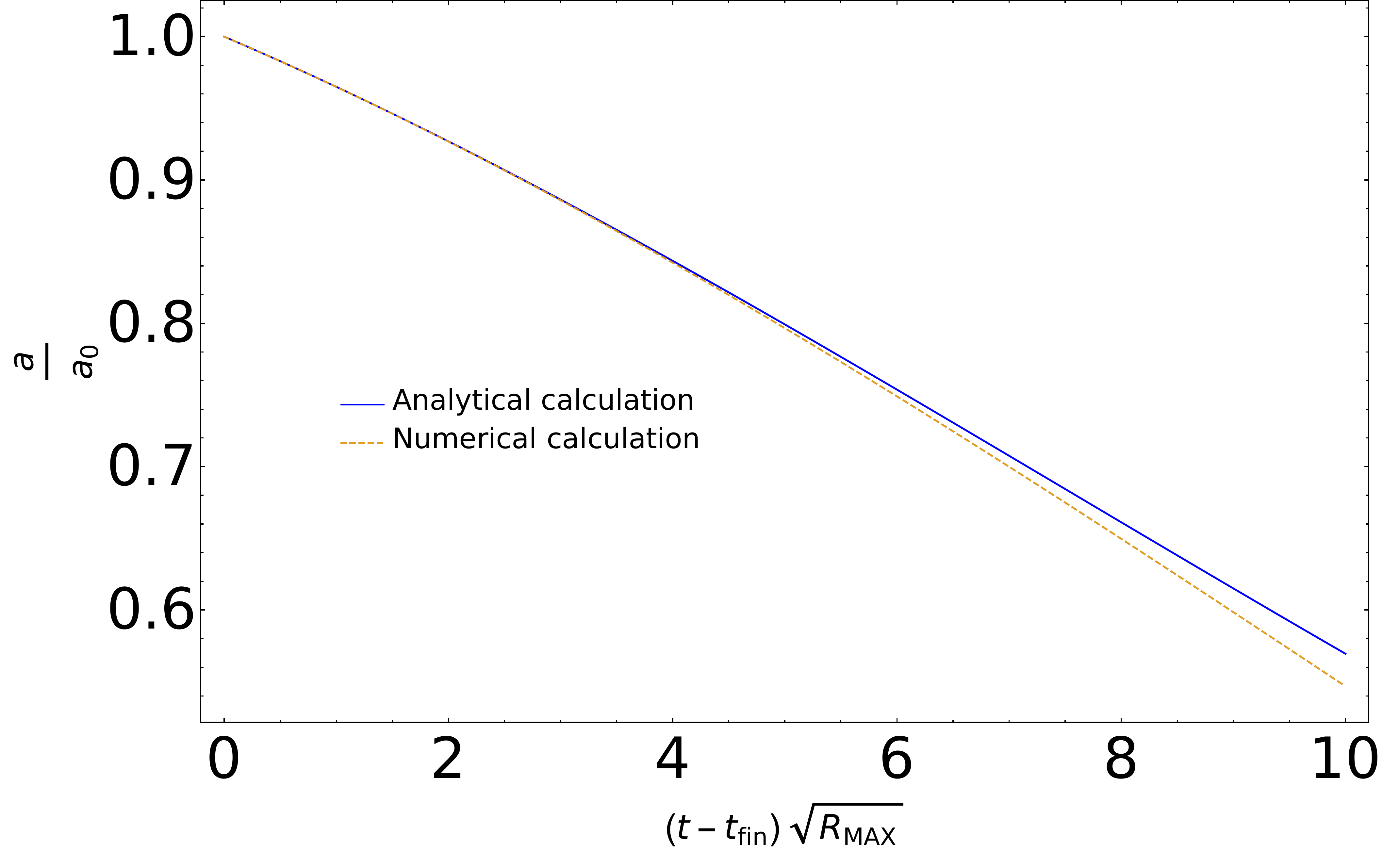}
  \caption{Time evolution of the scale factor, $a(t)$, during the transition to a 
  cosmological bounce following (\ref{hubl-skupljanje}), with parameters $A_{1}=-0.005$, $H_{fin}=-0.03$ with
  $N=1$.
   The numerical result (dashed line) is obtained by solving (\ref{frid-skupljanje}) with $\Lambda=0.01$, $A_{1}=-0.005$, $H_{fin}=-0.03$, 
   using (\ref{par4}) and (\ref{par5}).}
  \label{skalaslikaanti}
  \end{figure}

When the second order term again becomes significant, modified Friedmann equation must lead to solutions which will enable
the Universe to contract with an increasing curvature scalar, in order to reach a new bouncing phase and start a new cycle, thus $\dot{R}(t)>0$, $\dot{a}(t)<0$.
As was the case with the early expansion phase, many different solutions will enable this transition.
In general, the contracting Universe in the Einstein-Hilbert low curvature regime which enters into contracting
phase with a small cosmological constant will continue
to contract, therefore again leading to high curvature 
regime and the bouncing phase.
 But  for the sake of concreteness let us concentrate
on a particularly simple case where the (negative) Hubble parameter has a polynomial dependence on time
\begin{equation}
H(t)=\sum_{i=1}^{N}A_{i}(t-t_{fin})^{i} + H_{fin}, 
\label{hubl-skupljanje}
\end{equation}
where $t_{fin}$ is the time when the rip phase ends, and obviously $H(t_{fin})<0$. Since the rip phase ends with a negative
value of the Ricci scalar there is a further constraint on parameters: $A_{1}< -2 H(t_{fin})^2$. 
Since we are again dealing with the second order curvature correction
to standard GR Lagrangian, the considered $f(R)$ has the same form as in (\ref{early}), but now
with a different value for the parameters $c_{1}$ and $c_{2}$. Using the modified Freidman equation (\ref{numbounce}), with 
the curvature expansion to the second order, and taking the Universe to still be essentially empty while ignoring the 
variation of $\Lambda$, the modified Friedmann equation in this phase reads
\begin{multline}
3H^{2}(t)[c_{1}+ 2c_{2}(6 \dot{H}(t)+12H^{2}(t))]=\frac{c_{2}}{2}[6\dot{H}(t) + 12H^{2}(t)]^{2} \\
+\Lambda - 6 H(t) c_{2}[6\ddot{H}(t)+ 24 \dot{H}(t) H(t)].
\label{frid-skupljanje}
\end{multline}
Solving (\ref{frid-skupljanje}) with the ansatz (\ref{hubl-skupljanje}) in the simplest case, $N=1$, we obtain the solutions
\begin{equation}
 c_{1}=\frac{2\Lambda}{A_{1}},
 \label{par4}
\end{equation}
\begin{equation}
 c_{2}=-\frac{\Lambda}{18A_{1}}.
 \label{par5}
\end{equation}
We depict $H(t)$, $R(t)$ and $a(t)$ corresponding to (\ref{hubl-skupljanje}) and compare it with numerical results obtained by solving (\ref{frid-skupljanje})
using (\ref{par4} - \ref{par5}) in Fig. \ref{hubbleslikaanti}, \ref{riccislikaanti} and \ref{skalaslikaanti}. 
\\
\\
When the value of the Ricci scalar
increases enough, higher order terms in the field Lagrangian will become relevant, leading to a new bounce - as already described
in section 1. In this way the Universe eternally oscillates between bouncing phases, undergoing all phases of the standard
$\Lambda$CDM cosmology in the low curvature limit, where the effective corrections coming from the quantum effects can be ignored. Due to the variation
of the expansion factors of the Lagrangian, the standard $\Lambda$CDM phase is followed by a cosmological rip and the beginning of a contraction phase, opening
the way for the beginning of a new cycle in the infinite history of the Universe. 
\clearpage

\section{Examples of concrete models supporting the cyclic cosmology}
\begin{figure}[t]
  \centering
  \includegraphics[scale=0.23]{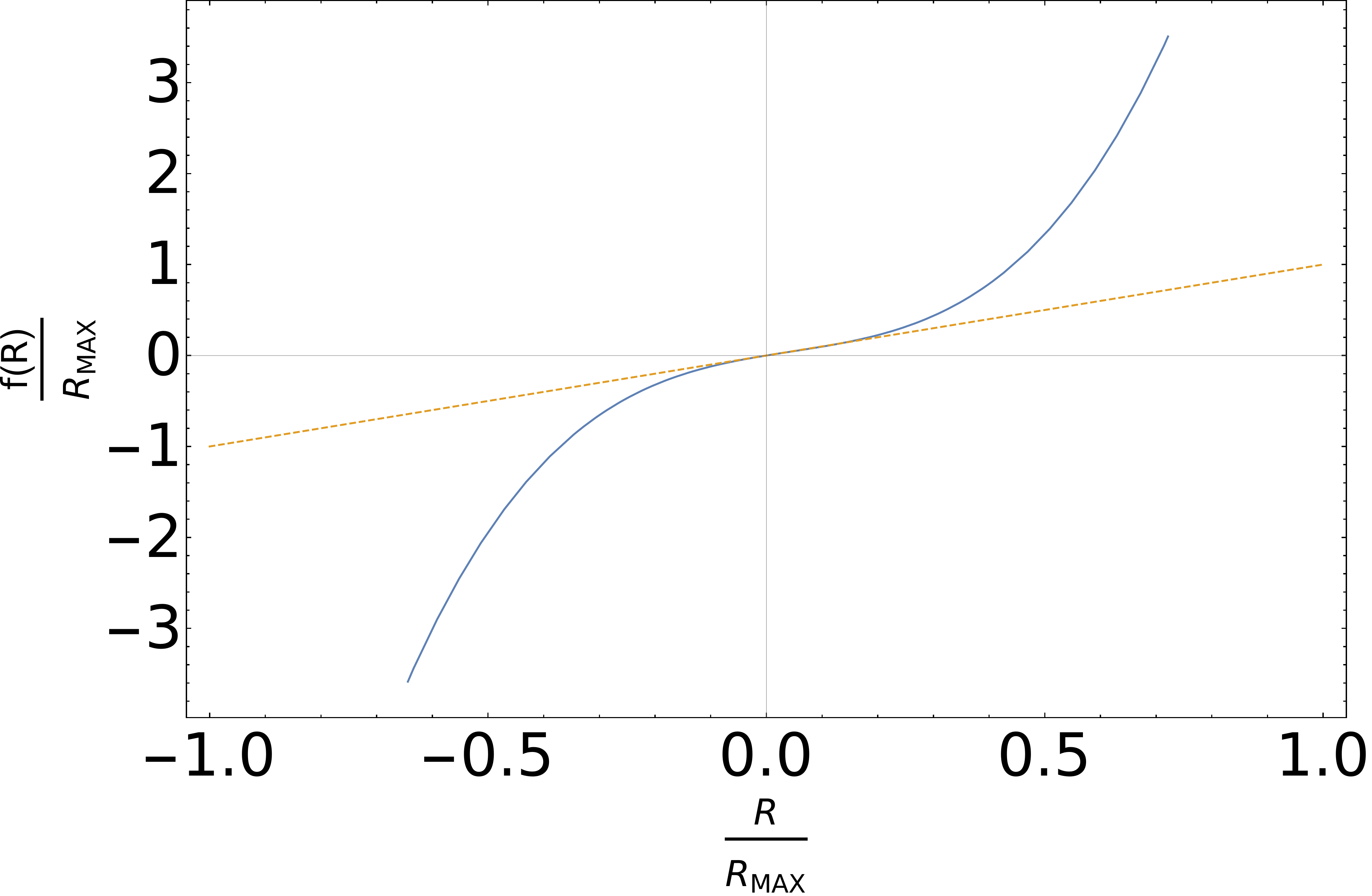}
  \caption{Typical form of $f(R)$ function leading to the bouncing solution in 
the third order expansion, given by Eq. (\ref{suma}), (\ref{par2}) and (\ref{par3}) (full line), 
compared to the standard GR with cosmological constant (dashed line). 
   Here $\Lambda = 0.0005$, $R_{2} = -2$.}
  \label{rek}
  \end{figure}

\begin{figure}[t]
  \centering
  \includegraphics[scale=0.23]{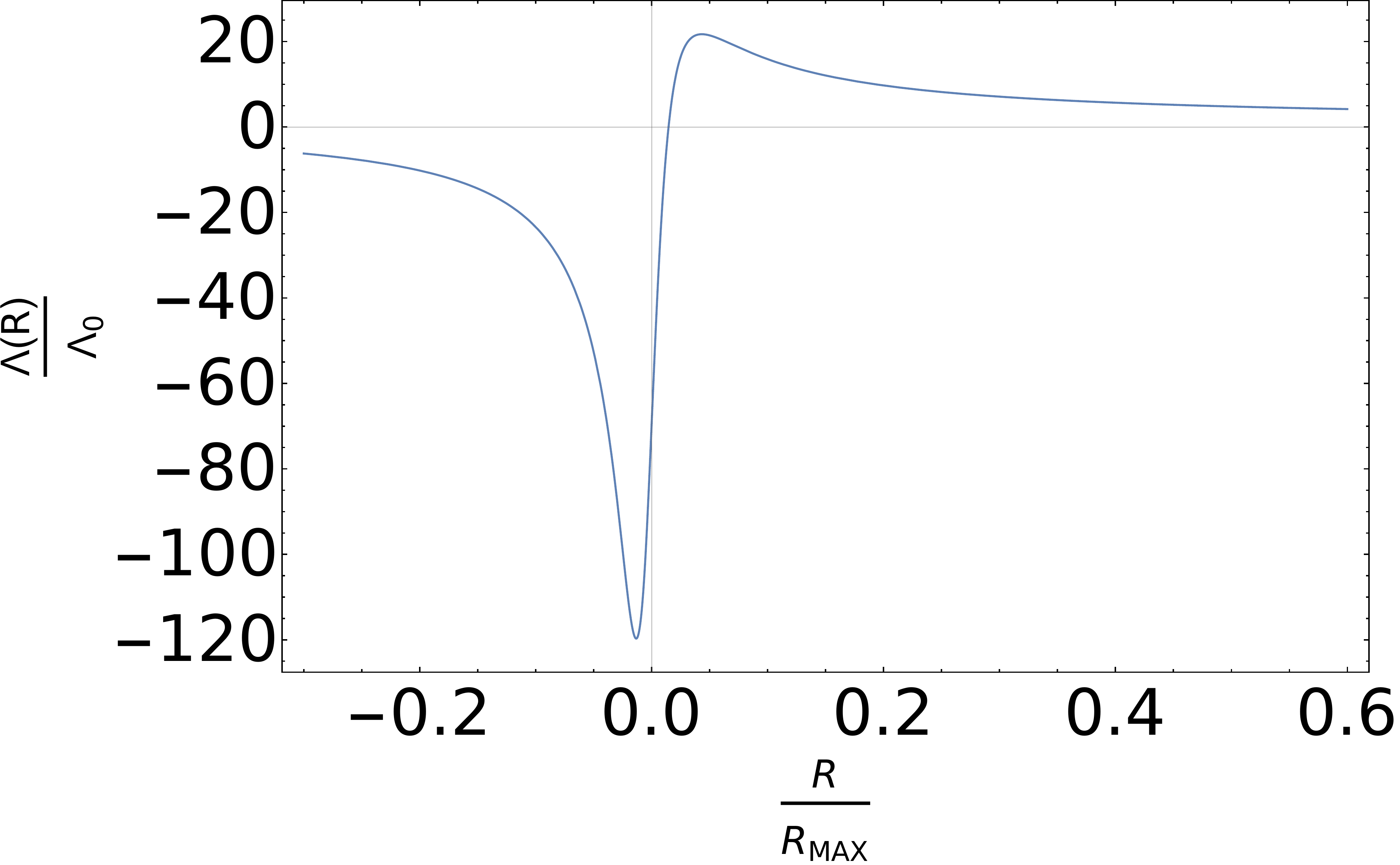}
  \caption{Typical scaling of cosmological term, $\Lambda(R) $, with curvature, 
as discussed in section VI, 
and given by Eq. (\ref{reklamb}). 
   Here $R_{min} = -0.5/R_{max}$, $B = 2/ R_{max}$, $A$ = 0.5}
  \label{reklambd}
  \end{figure}

As previously discussed, it is natural to assume that quantum corrections to the Lagrangian density of GR will have a different form in 
different curvature regimes. It is therefore not probable that the effects of quantum gravity could be fully modeled with a single $f(R)$ function. 
In accord with this reasoning, we based the proposed model of cyclic cosmology
on model independent analysis - deriving specific conditions and relations between coefficients of curvature corrections specific for each regime of 
the cyclic universe. However, it is interesting to
ask what would be the archetypal forms of Lagrangian modifications supporting the cyclic cosmology, and therefore in this section we reconstruct 
concrete example of $f(R)$ function leading
to cosmological bounce, rip and contraction, based on our previous discussions. \\ \\
Using the equation (\ref{suma}) to the third order, as well as (\ref{par2}) and (\ref{par3}) and prescribing $\Lambda$ and $R_{2}$ we can obtain a specific model consistent 
with bouncing solutions, which we depict 
on Fig. (\ref{rek}). 
Here we have assumed that variation of coefficients of the expansion can be neglected during the bouncing phase. As expected, we see that the
action integral for gravity 
effectively reduces to the GR case when curvatures are small compared to the maximal curvature of the Universe, reached at the bouncing phase. To couple 
this regime with the later rip phase
and turnaround, we need to take into account the effective scaling of cosmological term, $\Lambda(R)$, as proposed in the section VI. As a simple test 
function that satisfies necessary
properties described in section VI. we take 
\begin{equation}
\Lambda(t)=\Lambda_{0}- A \frac{1-(t-t_{max})^2}{1+(t-t_{max})^4},
\end{equation} 
where $t_{max}$ is the time when the 
scale factor reaches its maximum, and $A$ is
a constant. 
Since the Ricci scalar has a minimum in this phase it can be approximated as $R(t)=R_{min} +B(t-t_{max})^{2}$, which is also consistent with the solution of 
the equation (\ref{ripanje}) around zero point
of $\Lambda(t)$, given by: $H(t)=C(t-t_{max})$. We therefore obtain the following dependence of cosmological term on the curvature
  
\begin{equation}
 \Lambda(R)=\Lambda_{0}- A \frac{1-(R-R_{min})/B}{1+(R-R_{min})^2/B^2},
 \label{reklamb}
\end{equation}
which we show in Fig. (\ref{reklambd}).

\begin{figure}[t]
  \centering
  \includegraphics[scale=0.23]{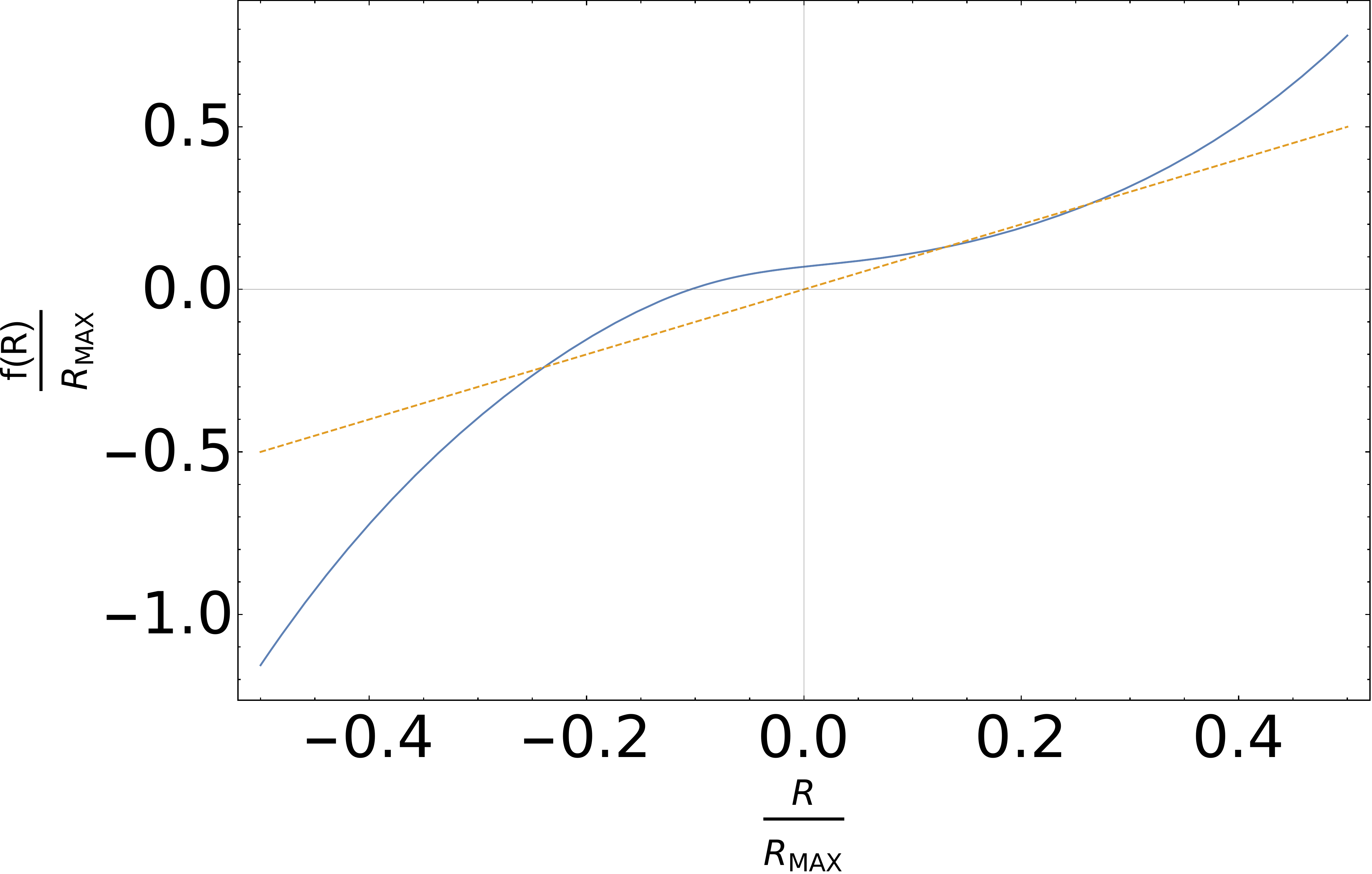}
  \caption{   Typical form of $f(R)$ function leading to cosmological bounce, 
    rip phase and entering into contracting phase (full line), 
    obtained by combining third order expansion of the action, 
    given by Eq. (\ref{suma}), (\ref{par2}) and (\ref{par3}), 
together with the scaling of $\Lambda (R)$ given by equation (\ref{reklamb}) 
compared to the standard GR with cosmological constant (dashed line). 
   Here $\Lambda = 0.0005$, $R_{2} = -2$, $R_{min} = -0.5/R_{max}$, $B = 
 2/ R_{max}$, $A$ = 0.5.}
  \label{ukup}
  \end{figure}

Finally, neglecting the variation of all other coefficients, we show the complete modified Lagrangian 
density, $f(R)=-2\Lambda(R) + (R/R_{max})+ c_{2}(R/R_{max})^{2}
+ c_{3}(R/R_{max})^{3}$, leading to bouncing, rip and contraction phase in Fig (\ref{ukup}). 
When the values of the Ricci curvature scalar are 
high compared to $R_{max}$, $\Lambda$ can be considered
as constant, but higher order corrections cause a significant departure from the GR description. This corresponds to the bouncing and 
early expansion phase. Then, as curvature decreases
during the evolution of the universe, contribution of the higher order terms become negligible, and the action approximately matches 
the one corresponding to the standard GR with the constant
cosmological term, leading to $\Lambda$CDM cosmological phase. Finally, at the end of asymptotic De-Sitter phase, $\Lambda$ starts to 
vary significantly. Its value rises at first, leading to the
rapid expansion and ripping of all bounded systems, and then abruptly decreases becoming negative - leading to decelerated expansion of 
the universe and finally beginning of the contracting phase. Then subsequently, the increase of the curvature 
scalar during the contraction phase causes the higher 
order corrections to again become significant leading to
the new bouncing phase.

\section{Conclusion}
  
  The question of the origin of the Universe, and its later evolution, has always been one of the leading intellectual driving forces in the development
  of mythologies, religions, philosophical systems and physical theories. In the last hundred years the development of the physical cosmology has 
  enabled us to make the first steps in a quantitative and empirical understanding of this question. The standard cosmological $\Lambda$CDM model based
  on General Relativity, and the \textit{ad hoc} addition of dark matter and dark energy, predicts the beginning of the Universe in the initial singularity.
  But, one must admit that up to this date some of the crucial assumptions of the $\Lambda$CDM model have not been empirically verified.
  Moreover, singularity theorems which lead to the initial singularity are dependent on the validity of Einstein's GR, which cannot be a complete theory
  since it does not take into account the principles of quantum physics. Taking the usually assumed position that the quantization of gravity will remove  
  singularities existing in GR, it seems natural to replace the Big Bang paradigm with the idea of an eternally existing cyclic Universe.
  \\
  \\
  Inspired by the works on cyclic cosmology in the previous decades, we propose a new potentially viable model of cyclic cosmology which is assuming 
  no hypothetical ingredients such as extra dimensions, new scalar fields, phantom energy and special space-time geometries.  
  Also, we do not use any specific theoretical framework, such as string theory or loop quantum gravity. 
  In our approach we start from the idea that quantum gravity effects can be modeled by
  higher order curvature corrections to the standard GR Lagrangian density for gravity.
  For the sake of simplicity we describe these contributions in the framework of $f(R)$ in the metric formalism, using a power law expansion
  in terms of the Ricci curvature scalar.
  We also take into consideration that a given power expansions, corresponding to a different $f(R)$ function, will in general be dependent on 
  a specific curvature regime. In this model 
  each cycle of the eternally existing Universe starts from the cosmological bounce, which ends the contraction
  phase of a previous cycle, gradually leading to the regime of standard $\Lambda$CDM cosmology. At the end of this regime the variation of the coefficients 
  leads to a non-singular rip of bounded systems, where the non-Hilbert terms contribution plays a dominant role and leads to the transition to a contraction 
  phase of the Universe.
  Then, during the contraction of an essentially empty Universe, higher order curvature terms again become significant and lead to a new bounce 
  and the beginning of a new cycle. 
  \\
  \\
  We have first analyzed the mathematical properties and necessary conditions for the establishment of a bounce - a transition from contraction to an expanding phase
  of the Universe. It has been shown that a bouncing solution appears as a rather natural and general feature of the FLRW geometry. Namely, if the Ricci scalar has 
  a maximum at a certain point in time, then on some small interval around that point it is given by a function satisfying the requirements for a bouncing
  solution. The critical question is if these conditions can be satisfied in the framework of the given field equations. Using the perturbative approach up to
  the fourth order in time and third order in Ricci power law series, we have determined the relations between the coefficients of expansion and obtained the solutions 
  for $H(t)$ and $R(t)$, which we compare with numerical solutions -- showing a good agreement of different techniques. 
  Inflationary expansion of the Universe may be conceptually unnecessary in the framework of the cyclic Universe, but we still briefly discuss it in the following
  chapter for completeness, where we have shown that inflation can easily be obtained in our model. 
  After this potential phase of the early expansion - possibly incorporating the inflation phase - during te subsequent evolution of the
  Universe all higher order curvature corrections to GR Lagrangian density become negligible and its dynamics is approximately
  described by the standard $\Lambda$CDM model. The late time evolution of this regime leads to an essentially empty Universe dominated by the
  effective cosmological term. In our model the slow variation of the coefficients in the curvature corrections then leads to the strong growth of 
  terms playing the role of the effective dark energy, and therefore leading to the destruction of all bounded systems in the Universe - 
  in order to avoid problems with the increase of entropy during the contraction phase. We have then assumed that after this rip phase, the Universe
  reaches the maximal value of the scale factor, and starts the contraction. The equations leading to this scenario, as
  well as the necessary conditions for the rip and entering into the contraction period have then been studied. Absorbing all the time dependence of the expansion coefficients
  in the zeroth order term, and choosing a family of functions with a suitable asymptotical behavior, we also presented the corresponding solutions
  of the modified Friedmann equations. 
  To get the full period of a cycle we need one last phase - the contraction phase. In this regime, the Ricci scalar must increase in order to reach its maximum at a bounce,
  and in the same time the scale factor must decrease and reach its minimum at a bounce. Many different solutions will enable this transition, but for simplicity we demonstrated the features of this phase by 
  modeling the Hubble parameter as a decreasing polynomial function in time. We have then analytically solved the modified Friedmann equation to the third order in non-Hilbert
  curvature corrections. Again, we compared
  the numerical solution, using the same parameters obtained by the analytical method, with the simple analytical solution, and conclude that the solutions 
  in both techniques are in a good agreement. \\ \\
  We have thus analysed the basic aspects of a new general cyclic model based on the corrections and generalization of the standard GR. Further 
  work should concentrate on the question of potential observational consequences of this model, and also on the more detailed discussion of the open
  cosmological problems within this framework. In the present work it was assumed that the contribution of stress-energy tensor components
  can be neglected in all phases of the cycle, apart from the low curvature regime. It would be important to avoid this assumption and also to address
  in a greater detail the issue of entropy evolution during a cycle of the Universe, specially during the contracting phase. Moreover, it would also be interesting to
  apply the same scenario to the other variants of modified gravity formalisms, such as $f(T)$ gravity and Palatini or metric-affine $f(R)$ gravity formalism,
  to see if other formalisms lead to the same physical conclusions. \\
  \clearpage
  \begin{center}
   \textit{''This world, the same for all, neither any of the gods nor any man\\
has made, but it always was, and is, and shall be, an ever living\\
fire, kindled in due measure, and in due measure extinguished.”} \\
Heraclitus of Ephesus\cite{heraklid}
  \end{center}

\section*{Acknowledgement}
Authors would like to thank Natacha Leite for reading the manuscript and given suggestions.


\begin{thebibliography}{99}
\bibitem{ein1}
A. Einstein , ‘Die Grundlage der allgemeinen Relativitätstheorie’, Annalen der Physik.
49(7), 769-822.[1916a],Reprinted as Vol. 6, Doc. 30 CPAE, 
\bibitem{ein2}
A. Einstein, Albert, Die Feldgleichungen der Gravitation, Sitzungsberichte der Preussischen Akademie der Wissenschaften zu Berlin, (November 25, 1915) 844847.
\bibitem{berti}
E. Berti et al., 
Class. Quantum Grav. 32, 243001 (2015)
\bibitem{will}
C. M. Will,
Living Rev. Relativity 17 (2014), 4
\bibitem{tur}
S. G. Turyshev,
Ann.Rev.Nucl.Part.Sci.58:207-248 (2008)
\bibitem{ligo}
The LIGO Scientific Collaboration, the Virgo Collaboration,
Phys. Rev. Lett. 116, 221101 (2016)
\bibitem{ein3}
A. Einstein, [1917], ‘Kosmologische Betrachtungen zur allgemeinen Relativitätstheorie’, Sitzungs-.
berichte der Königlich Preussischen Akademie der Wissenschaften. Reprinted as Document 43.
(page 541) of Volume 6 CPAE
\bibitem{hu1}
E. Hubble, Proc. Natl. Acad., Sci., 15:168-173, (1929)
\bibitem{hu2}
 E. Hubble and M.L. Humason, 
Astrophys. J., 74:43–80, (1931)
\bibitem{fri}
A. Friedman, Uber die Krummung des Raumes, Zeitschrift fur Physik, 10, 377 (1922)
\bibitem{robertson}
H. P. Robertson, 
Reviews of Modern Physics, 5:62-90, 1933
\bibitem{weinberg}
S. W. Weinberg, Gravitation and Cosmology, Wiley, New York, 1972.
\bibitem{maartens}
 Ellis, G. F. R., Maartens, R., and MacCallum, M. A. H. 2012.
Relativistic cosmology. Cambridge: Cambridge University Press
\bibitem{perlmutter}
S.Perlmutter, et al.
Astrophys.J. 517, 565 (1999)
\bibitem{riess}
Riess et al. (Supernova Search Team Collaboration), Astron. J. 116 (1998) 1009, astro-ph/9805201.
\bibitem{eisenstein}
D. J. Eisenstein et al. (SDSS Collaboration), Astrophys. J. 633 (2005), 560, astroph/0501171.
\bibitem{spergel}
D. N. Spergel et al. (WMAP Collaboration), Astrophys. J. Suppl. 148 (2003), 175, astroph/0302209.
\bibitem{ost}
J. P. Ostriker, P. J. E. Peebles, ApJ 186, 467-480 (1973).
\bibitem{refre}
A. Refregier, Ann.Rev.Astron.Astrophys.41:645-668, (2003).
\bibitem{mas}
 R. Massey, J. Rhodes, R. Ellis,
et al., Nature 445 286-290 (2007).
\bibitem{ol}
K.  A.  Olive et  al.
(Particle  Data  Group),  Chin.  Phys.  C,  38 090001 (2014) 
\bibitem{planco}
 Planck Collaboration, P. A. R. Ade, N. Aghanim, C. Armitage-Caplan, et al., Astron.Astrophys 571 A16 (2014)
 \bibitem{bull}
 Bull, Philip et al. Phys.Dark Univ. 12 (2016) 56-99
\bibitem{carroll}
Carroll, S. M., Living Rev. Rel.4, 1 (2001).
\bibitem{wein}
Weinberg, S., Rev. Mod. Phys.
61, 1 (1989)
\bibitem{timothy}
 T. Clifton, P. G. Ferreira, A. Padilla, C. Skordis,
Physics Reports 513, 1 (2012), 1-189
\bibitem{noj-odin}
S. Nojiri, S. D. Odintsov,
Phys.Rept. 505, 59-144 (2011)
\bibitem{brans}
C. Brans, R. H. Dicke
Phys. Rev. 124, 925 (1961)
\bibitem{cai}
Y. Cai, S. Capozziello, M. De Laurentis, E. N. Saridakis,
Rept.Prog.Phys. 79 no.4, 106901 (2016)
\bibitem{overduin}
J. M. Overduin, P. S. Wesson,
Phys.Rept.283:303-380 (1997)
\bibitem{thom}
T. P. Sotiriou,
J.Phys.Conf.Ser.283:012034 (2011)
\bibitem{faraoni}
 T. P. Sotiriou, V. Faraoni,
  Rev. Mod. Phys. 82 451-497 (2010)
\bibitem{stelle}
Stelle, K. S.,
Phys. Rev. D16, 953. (1977)
\bibitem{vilkovisky}
Vilkovisky, G. A., 
Class. Quant. Grav. 9, 895 (1992)
\bibitem{haw1}
S. W. Hawking. Roy. Soc. P c. A-Math Phy., 294(1439):511- 521. (1966 )
\bibitem{haw2}
S. W. Hawking. Roy. Soc. II. P c. A-Math Phy., 295(1443):490-493. (1966)
\bibitem{haw3}
S. W. Hawking. Roy. Soc. III. P c. A-Math Phy., 300(1461):187-201. (1967)
\bibitem{linde}
A. Linde, Lect.Notes Phys. 738: 1-54 (2008)
\bibitem{bojo}
M. Bojowald, AIPConf.Proc. 910:294-333 (2007)

\bibitem{tol}
R. C. Tolman, Phys. Rev 38, 1758 (1931)



\bibitem{lem}
G. Lemaitre, Annales Soc. Sci. Brux. Ser. I A
53, 51 (1933).
\bibitem{lifshitz}
 E. M. Lifshitz, I. M. Khalatnikov, Adv. Phys. 12, 185-249 (1963). 
\bibitem{misner}
C. W. Misner, Phys. Rev. Lett. 22, 20 (1969)
\bibitem{dicke}
R. Dicke, P. J. E. Peebles, General relativity: An Einstein Centenary Survey, 
ed. by S. Hawking and W. Israel (Cambridge Univ. Press, 1979) 
\bibitem{markov}
M. A. Markov,
Annals of Physics 155, 333-357 (1984)
\bibitem{cai2}
Y.F. Cai, E. N. Saridakis,
Journal of Cosmology, Volume 17, 7238-7254 (2011)
\bibitem{ivanov}
R. I. Ivanov, E. M. Prodanov
Phys. Rev. D 86, 083536 (2012)
\bibitem{salah}
M. Salah, F. Hammad, M. Faizal, A. F. Ali,
arXiv:1608.00560 [gr-qc]
\bibitem{varun}
V. Sahni, A. Toporensky,
Phys. Rev. D 85, 123542 (2012)
\bibitem{singh}
P. Singh, K. Vandersloot, G. V. Vereshchagin
Phys. Rev. D 74, 043510 (2006)
\bibitem{biljar}
A. P. Billyard, A. A. Coley, J. E. Lidsey,
 J.Math.Phys. 41 6277-6283 (2000) 
\bibitem{sari}
Y. F. Cai, E. N. Saridakis,
Class. Quantum Grav. 28 (2011) 035010
\bibitem{amani}
A. R. Amani, Int. J. Mod. Phys. D, 25, 1650071 (2016)
\bibitem{lopez}
M. Bouhmadi-Lopez , J. Morais, A. B. Henriques,
Phys. Rev. D 87, 103528 (2013)
\bibitem{stach}
T. Stachowiak, M. Szydlowski
Phys. Lett. B646:209-214 (2007)
\bibitem{moriconi}
R. Moriconi, G. Montani, S. Capozziello,
Phys. Rev. D 94, 023519 (2016)
\bibitem{visser}
C. Cattoen, Matt Visser,
Class.Quant.Grav. 22 4913-4930 (2005) 
\bibitem{roshan}
M. Roshan, F. Shojai,
Physical Review D 94, 044002 (2016) 
\bibitem{noj-odin2}
S. D. Odintsov, V. K. Oikonomou,
Phys. Rev. D 91, 064036 (2015)
\bibitem{cai3}
Y. F. Cai, S. H. Chen, J. B. Dent, S. Dutta and E. N. Saridakis, Class. Quantum Grav. 28, 215011 (2011)
\bibitem{gurovich}
V. Ts. Gurovich, Soviet Physics Doklady, Vol. 15, p.1105 (1971)
\bibitem{ruter}
J.C. Fabris, S. Reuter, GRG 32, 1345 (2000)
\bibitem{khoury}
J. Khoury, B. A. Ovrut, P. J. Steinhardt, N. Turok,
Phys. Rev. D64, 123522 (2001)
\bibitem{doom}
R. R. Caldwell, M. Kamionkowski, N. N. Weinberg
Phys. Rev. Lett. 91, 071301 (2003)
\bibitem{carlson}
E. D. Carlson, P. R. Anderson, J. R. Einhorn, B. Hicks, A. J. Lundeen,
arXiv:1607.01699 [gr-qc]
\bibitem{davies}
D. Davies, Annales Poincare Phys. Theor. 49, 297 (1988)
\bibitem{brown}
M. G. Brown, K. Freese, W. H. Kinney
JCAP 0803:002 (2008)
\bibitem{lehners}
Lehners, Jean-Luc Phys.Rept. 465 (2008)
\bibitem{steinhardt}
P. J. Steinhardt, N. Turok, Science 296, 1436 (2002)
\bibitem{george}
G. F. R. Ellis, E. Platts, D. Sloan, A. Weltman, JCAP 1604, no.04, 026 (2016) 
\bibitem{gomez}
 S.~Nojiri, S.~D.~Odintsov and D.~Saez-Gomez,
  AIP Conf.\ Proc.\  {\bf 1458}, 207 (2011)
\bibitem{piao1}
Y. S. Piao, B. Feng, X. Zhang,
Phys. Rev. D 69, 103520 (2004)
\bibitem{piao2}
Z. G. Liu, Z. K Guo, Y. S. Piao,
Phys. Rev. D 88, 063539 (2013)

\bibitem{landau}
  L. D. Landau and E. M. Lifshitz, The Classical Theory of Fields, 4th ed., Course of Theoretical Physics
(Butterworth-Heinemann, Oxford, 1975).
\bibitem{misner0}
C. W. Misner, K. S. Thorne, and J. A. Wheeler, Gravitation (W. H. Freeman, New York, 1973).
\bibitem{buc}
H. A. Buchdahl, Mon. Not. R. Astron. Soc., 150, 1 (1970).
\bibitem{cembranos}
J. A. R. Cembranos, Phys.Rev.Lett.102:141301, (2009).









\bibitem{tombolis}
E. Tomboulis, Phys. Lett. 70B (1977) 361; Phys. Lett. 97B (1980) 77;
\bibitem{steel}
K. Stelle, Phys. Rev.D16 (1977) 953
\bibitem{salam}
A. Salam and J. Strathdee, Phys. Rev.D18 (1978) 4480.
\bibitem{julve}
J. Julve and M. Tonin, Nuovo Cim. 46B (1978) 137.
\bibitem{motola}
B. Hasslacher and E. Mottola, Phys. Lett. 99B (1981) 221.
\bibitem{fedkin}
E. Fradkin and A. Tseytlin, Nucl. Phys. B201 (1982) 469; Phys. Lett. 104B (1981) 377.
\bibitem{eliza}
E. Elizalde, S. Odintsov and A. Romeo, Phys. Rev. D51 (1995) 1680, 4250.
\bibitem{psa}
 D. Psaltis, D. Perrodin, K. R. Dienes, and I. Mocioiu, Phys. Rev. Lett., 100, 091101 (2008).
 \bibitem{stabile}
 S. Capozziello, A. Stabile, and A. Troisi, Phys. Rev. D, 76, 104019 (2007).
 \bibitem{faulkner}
 T. Faulkner, M. Tegmark, E. F. Bunn, and Y. Mao, Phys. Rev. D, 76, 063505 (2007).
\bibitem{clif}
T. Clifton, Phys. Rev. D, 77, 024041 (2008). 
\bibitem{berry}
C.P.L. Berry, J. R. Gair, Phys.Rev.D83:104022, (2011).
\bibitem{sigurna1}
 R. Percacci, In *Oriti, D. (ed.): Approaches to quantum gravity* 111-128 arXiv:0709.3851 [hep-th]
 \bibitem{sigurna2}
M. Niedermaier, Class.Quant.Grav. 24 (2007) R171-230 gr-qc/0610018
\bibitem{elise}
 S. W. Hawking and G. F. R. Ellis, \textit{The Large Scale Structure of Sp
ace-Time}, (Vol. 1) Cambridge University Press 1973 
\bibitem{woodard}
R. P. Woodard
Lect.Notes Phys.720:403-433 (2007)
\bibitem{dolgov}
Dolgov, A. D., and M. Kawasaki, Phys. Lett.
B573, 1. (2003)
\bibitem{sawicki}
Sawicki, I.,W. Hu, Phys. Rev. D75, 127502 (2007)


  \bibitem{starob}
A. A. Starobinsky, Phys. Lett. B 91, 99 (1980)

\bibitem{novi}
E. D. Carlson, P. R. Anderson, J. R. Einhorn, B. Hicks, A. J. Lundeen, arXiv:1607.01699.
\bibitem{novi2}
L. Perivolaropoulos, Phys. Rev. D 94, 124018 (2016).
\bibitem{heraklid}
From The Fragments of the Work of Heraclitus of Ephesus on Nature, translated from the Greek text of Bywater by G.T.W. Patrick, Baltimore: N. Murray, 1889.


\end{thebibliography}
\end{document}